\documentclass[a4paper]{article}

\usepackage{
    physics, 
    graphicx,
    amsmath,
    amssymb,
    mathtools,
    amsfonts,
    caption,
    todonotes,
    subcaption,
    geometry,
    siunitx,
    setspace,
    booktabs
}

\usepackage[version=4]{mhchem}
\usepackage[final]{changes}
\newcommand{\add}{\color{black}}

\newcommand{\ts}[1]{_{\text{#1}}}
\newcommand{\kin}[1]{\text{k} \in \{\text{#1}\}}
\newcommand{\mybar}{\overline}

\begin{document}

\title{An asymptotic derivation of a single particle model with electrolyte}

\author{
    Scott G. Marquis 
        \footnote{\texttt{marquis@maths.ox.ac.uk}}
        \footnote{Mathematical Institute, University of Oxford, OX2 6GG, United Kingdom}
        \and
    Valentin Sulzer \footnotemark[2] \and
    Robert Timms \footnotemark[2] \footnote{The Faraday Institution} \and
    Colin P. Please \footnotemark[2] \footnotemark[3] \and
    S. Jon Chapman  \footnotemark[2] \footnotemark[3]
}

\maketitle

\begin{abstract}
    The standard model of a lithium-ion battery, the Doyle-Fuller-Newman (DFN) model, is computationally expensive to solve. Typically, simpler models, such as the single particle model (SPM), are used to provide insight. Recently, there has been a move to extend the SPM to include electrolyte effects to increase the accuracy and range of applicability. However, these extended models are derived in an ad-hoc manner, which leaves open the possibility that important terms may have been neglected so that these models are not as accurate as possible. In this paper, we provide a systematic asymptotic derivation of both the SPM and a correction term that accounts for the electrolyte behavior. Firstly, this allows us to quantify the error in the reduced model in terms of ratios of key parameters, from which the range of applicable operating conditions can be determined. Secondly, by comparing our model with ad-hoc models from the literature, we show that existing models neglect a key set of terms. In particular, we make the crucial distinction between writing the terminal voltage in pointwise and electrode-averaged form, which allows us to gain additional accuracy over existing models whilst maintaining the same degree of computational complexity.
\end{abstract}

\section{Introduction} 
\subsection{Background}
Lithium-ion batteries are used extensively in consumer electronics and industry{\add, and as a result, understanding the physical processes occurring in such systems is key}. Mathematical models are an essential tool for the design and management of battery systems. The standard mathematical model of a lithium-ion battery is the Doyle-Fuller-Newman (DFN) model, which was developed by John Newman and collaborators \cite{doyle,Fuller1994,newman_book}. This model is sometimes referred to as the pseudo-two-dimensional (P2D) model or simply the Newman model. The model consists of a set of highly-coupled nonlinear parabolic and elliptic partial differential equations (PDEs). {\add In the literature,} this system of equations has been solved using a variety of different numerical methods including finite-difference methods, control volumes, finite-element methods, and orthogonal collocation, among others \cite{doyle, Bizeray2015, Cai2012a, zeng2013efficient, Methekar2011, Subramanian2009, Northrop2011a}. However, even when employing sophisticated numerical techniques, the DFN model remains too computationally complex for some applications. For instance, battery management systems (BMS) adopt equivalent circuit models, which consist of only a handful of ordinary differential equations (ODEs) and are favoured over the DFN model due to a combination of factors such as speed, memory requirements, and numerical convergence. Additionally, in the study of coupled systems of battery cells (e.g. thermal effects in a battery pack), spending computational resources (both RAM and CPU) on a model as detailed as the DFN model is often unnecessary. For these applications and others, physics-based models that are simpler than the DFN model are desired. The simplest of these, the SPM, has been employed in several settings in recent years \cite{Moura2017, bizeray2018identifiability, guo2011single, dey2014combined, jorn}. There has also been \deleted{several previous}{\add a number of} papers that provide justification for the SPM and suggest correction terms that may increase the accuracy of the \deleted{voltage prediction it gives}{\add predicted voltage} \cite{kemper2013extended, Perez2016, prada2012simplified, han2015simplification, rahimian2013extension, tanim2015temperature}. However, these approaches generally rely on a number of ad-hoc assumptions. In this work, we provide a systematic mathematical derivation of the SPM and an additional correction for the electrolyte by applying asymptotic methods to the DFN model. Asymptotic methods are widely applied within many subdisciplines of mathematics, but they are still relatively underutilized in battery modelling \cite{hinchperturbation, bender2013advanced}. However, they have been successfully applied in the derivation of the DFN model (by means of asymptotic homogenization) and in the derivation of reduced-order lead-acid battery models \cite{Richardson2011, sulzer2019faster}. Asymptotic methods have also been applied to the reduction of lithium-ion battery models that neglect the effects of the particles \cite{moyles2018asymptotic}. 

In our approach we consider approximations that can be found by exploiting two physically-relevant limits: i) that the electrical conductivity is large in the electrodes and electrolyte (such that the typical potential drop in each material is small relative to the thermal voltage), and ii) that the timescale of lithium-ion migration in the electrolyte is small relative to the typical timescale for a discharge. The validity of applying both of these limits is determined directly from the input parameter values which allows for the errors in the reduced models to be estimated a-priori. By comparison, in \cite{Moura2017} for example, six assumptions that can only be validated a-posteriori by comparison with the full DFN model are required (e.g. that the current profile assumes a specific form). A key result of this work is the derivation of a single additional partial differential equation (PDE) and algebraic correction to the terminal voltage that accounts for nonuniform effects in the electrolyte and greatly improves the accuracy of the predictions when compared to the SPM. Additionally, we identify a key step overlooked in the ad-hoc derivations and show, through direct numerical comparison, that performing this step allows our reduced model to outperform the other ad-hoc models in the literature whilst retaining the same level of computational complexity.

After completion of the work in this paper, we became aware of another in-progress paper that employs asymptotic methods to simplify the DFN model \cite{foster2019}. In \cite{foster2019}, the asymptotic limit of large changes in the open-circuit voltage (OCV) relative to the thermal voltage is taken; this is a different (and in fact complementary) limit to that taken here. Their limit recovers a variant of the SPM at leading order because the reaction overpotentials are small. In contrast, in our limit the small gradients in the electrolyte concentration, electrolyte potential, and  electrode potentials give rise to the homogeneous behavior of electrode particles and hence the SPM. Whilst both limits recover a variant of SPM at leading order, the correction terms are different, for example \cite{foster2019} requires a second-order ODE for the electrolyte potential to be solved numerically. This results in a model that is slightly more computationally expensive than the model presented here but is significantly less computationally expensive than the DFN model. Therefore, if the model parameters for a particular case are not appropriate for the limit considered here, we encourage the reader to consider the reduced model in \cite{foster2019}.

We begin by providing a brief overview of the DFN model, after which we nondimensionalize the model by introducing typical scalings. At this stage, we re-write the terminal voltage in electrode-averaged form, which is essential in the derivation of our reduced models. We then identify key dimensionless parameters and perform a systematic asymptotic reduction in the distinguished limit in which the electrical conductivities are of a comparable size to the ratio of the typical discharge timescale to the lithium-ion migration timescale. We make a uniform asymptotic expansion and, at leading order, recover the SPM. By extending the asymptotic expansion to first order, we obtain an additional PDE for the concentration of lithium ions in the electrolyte {\add and} an additional algebraic correction to the terminal voltage. We summarize our reduced model and compare its computational complexity and accuracy with the DFN model when the finite-volume discretization method is applied to both models. We also discuss the reduced computational complexity of our reduced model when alternative discretization methods are employed. {\add Following our discussion on computational complexity}, we compare our model with a selection of ad-hoc models from the literature, showing that {\add our model} best recovers the predictions of the DFN model. Finally, we provide a summary of the dimensional version of our model in Section~\ref{sec:dimensional-SPMe}, alongside a set of conditions which are to be satisfied for our model to be applicable and a set of a-priori error estimates.

\section{Doyle-Fuller-Newman (DFN) model}
\begin{figure}[h]
  \centering
    \includegraphics{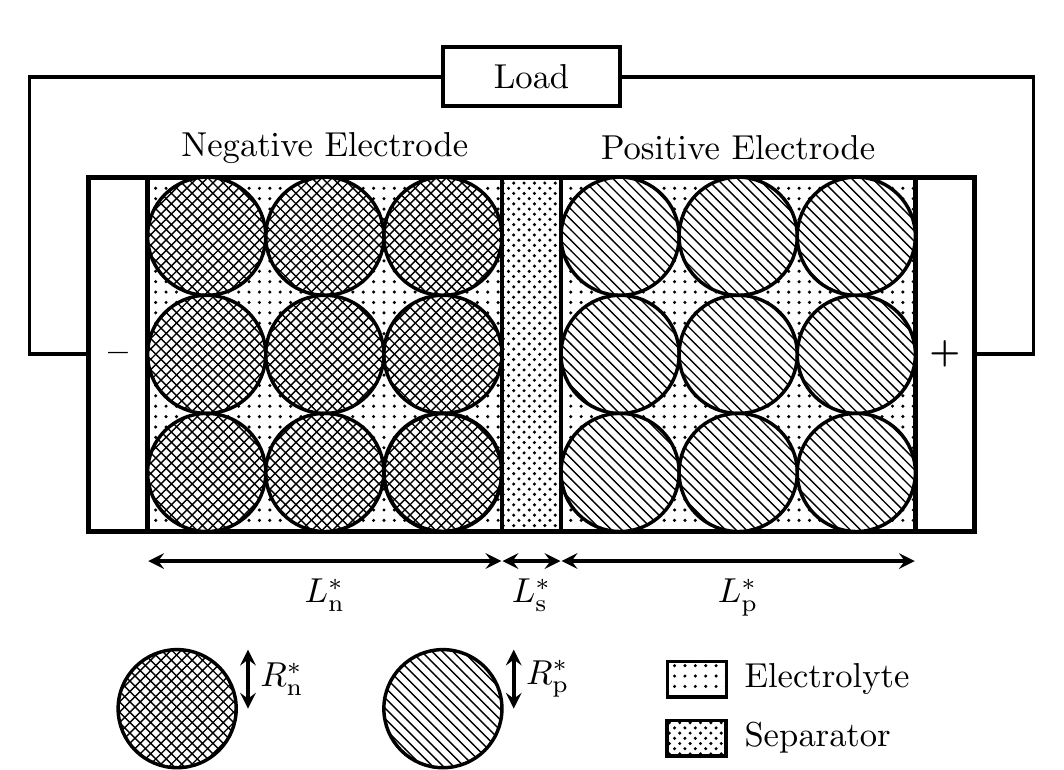}
  \caption{Schematic of a lithium-ion battery. Active material particles are shaded cross-stitch and diagonally for the negative and positive active materials, respectively.}\label{fig:liBat}
\end{figure}

Lithium-ion batteries consist of two electrodes, a porous separator, an electrolyte, and two current collectors, as displayed in Figure~\ref{fig:liBat}. Each electrode consists of active material particles within which lithium can be stored, and a binder {\add(not shown)} which holds the electrode together and maintains an electrical connection between the active material particles and the current collectors. \deleted{In Figure~\ref{fig:liBat}, we do not display the binder material. }

Upon discharge, lithium intercalated in the negative electrode particles diffuses to the surface of the particles where an  electrochemical reaction occurs. This electrochemical reaction produces a lithium ion free to move through the electrolyte and an electron free to move through the electrode. The electron travels through the electrode, into the current collector, through an external wire, and towards the positive electrode. Meanwhile, the lithium ion migrates through the electrolyte towards the positive electrode. At the surface of the positive electrode particles, the lithium ion and the electron combine through another electrochemical reaction to form a lithium atom intercalated in the positive electrode particle. To charge the battery, a voltage is applied across the cell and the whole process occurs in reverse.

Here, we summarize the DFN model, which is the standard model of a lithium-ion battery \cite{doyle,Fuller1994,newman_book}. The model is derived either by volume averaging \cite{newman_book} or the method of multiple scales \cite{Richardson2011}. Throughout, we use a superscript `$*$' to indicate dimensional quantities. As indicated in Figure \ref{fig:liBat}, the thicknesses of the negative electrode, separator, and positive electrode are $L^*\ts{n}$, $L\ts{s}^*$, and $L\ts{p}^*$, respectively. We denote the distance between the negative and positive current collectors by $L^*=L\ts{n}^*+L\ts{s}^*+L\ts{p}^*$. The active material particles in the negative and positive electrodes are assumed to be spheres with radii $R\ts{n}^*$ and $R\ts{p}^*$, respectively. Additionally, we assume that the behavior within each particle is spherically symmetric. We use the spatial coordinate $x^*\in[0,L^*]$ to indicate the location through the thickness of the battery and the spatial coordinate $r^*\in[0,R\ts{k}^*]$, $\kin{n, p}$ to indicate the location within each particle of active material. We define the following regions of the battery,
\begin{align*}
    \Omega^*\ts{n} = [0, L\ts{n}^*],             \quad
    \Omega^*\ts{s} = [L\ts{n}^*, L^*-L\ts{p}^*], \quad
    \Omega^*\ts{p} = [L^*-L\ts{p}^*, L^*],
\end{align*}
which correspond to the negative electrode, separator, and positive electrode regions, respectively. We denote electric potentials by $\phi^*$, current densities by $i^*$, lithium concentrations by $c^*$ (in the electrolyte $c^*$ refers to lithium-ion concentrations), and molar fluxes by $N^*$. To indicate the region within which each variable is defined, we include a subscript $\kin{n, s, p}$ which corresponds to $\Omega^*\ts{n}$, $\Omega^*\ts{s}$, and $\Omega^*\ts{p}$, respectively. To distinguish variables in the electrolyte from those in the solid phase of the electrode, we employ an additional subscript `$\text{e}$' for electrolyte variables and an additional subscript `$\text{s}$' for solid-phase variables. For clarity, the variables in the model and their corresponding regions are
\begin{align*}
   & \phi\ts{s,n}^*, \ \phi\ts{e,n}^*, \ c\ts{e,n}^*, \ i\ts{e,n}^*, \ N\ts{e,n}^*: \quad & x^*\in \Omega\ts{n}^*,                         \\
   & \phi\ts{e,s}^*, \ c\ts{e,s}^*, \ i\ts{e,s}^*, \ N\ts{e,s}^*: \quad                      & x^*\in \Omega\ts{s}^*,                         \\
   & \phi\ts{s,p}^*, \ \phi\ts{e,p}^*, \ c\ts{e,p}^*, \ i\ts{e,p}^*, N\ts{e,p}^*: \quad   & x^*\in \Omega\ts{p}^*,                         \\
   & c\ts{s,n}^*:                                                                      & r^*\in[0, R\ts{n}^*], \quad x^*\in \Omega\ts{n}^*, \\
   & c\ts{s,p}^*:                                                                      & r^*\in[0, R\ts{p}^*], \quad x^*\in \Omega\ts{p}^*.
\end{align*}
We note that $c\ts{s,n}^*$ and $c\ts{s,p}^*$ depend on the macroscopic spatial variable, $x^*$, the microscopic spatial variable, $r^*$, and time, $t^*$, whereas all other variables depend on $x^*$ and $t^*$ only. When stating the governing equations, we take the region in which an equation holds to be implicitly defined by the subscript, $\kin{n, s, p}$, of the variables. With this in mind, the DFN model is summarized as:
\begin{subequations}\label{eqn:DFN}
	\begin{align}
		\intertext{\textbf{Governing equations}}
        \intertext{\emph{Charge conservation:}}
		&\pdv{i\ts{e,k}^*}{x^*} = 
			\begin{cases} 
			  a\ts{k}^* j\ts{k}^*, \quad &\text{k} = \text{n, p}, \\
			  0, \quad &\text{k} = \text{s}, 
            \end{cases} 
		&& \kin{n, s, p},
		\\
		&i\ts{e,k}^* = \epsilon\ts{k}^b \kappa\ts{e}^*(c\ts{e,k}^*) 
			\left( 
				- \pdv{\phi\ts{e,k}^*}{x^*} 
				+ 2(1-t^+)\frac{R^*T^*}{F^*} \pdv{x^*}
					\left(
						\log(c\ts{e,k}^*)
					\right)
			\right), 
		&& \kin{n, s, p}, 
		\\ 
		\label{eqn:dim:solid_ohms} 
		&I^* - i\ts{e,k}^* = 
	  -\sigma\ts{k}^* \pdv{\phi\ts{s,k}^*}{x^*}, && \kin{n, p}.
        \intertext{\emph{Molar conservation:}}
		 &\epsilon\ts{k} \pdv{c\ts{e,k}^*}{t^*} 
			= \pdv{N\ts{e,k}^*}{x^*} 
			+ \frac{1}{F^*} \pdv{i\ts{e,k}^*}{x^*}, 
		 && \kin{n, s, p}, 
		\\
        &N\ts{e,k}^* 
		= \epsilon\ts{k}^b D_e^*(c\ts{e,k}^*) 
			\pdv{c\ts{e,k}^*}{x^*} + \frac{t^+}{F^*}i\ts{e,k}^*, 
		&&\kin{n, s, p},
		\\
		&\pdv{c\ts{s,k}^*}{t^*} = 
			\frac{1}{(r^*)^2} \pdv{r^*}
			\left( 
			  (r^*)^2 D\ts{s,k}^*\pdv{c\ts{s,k}^*}{r^*}
			\right)
	  , &&\kin{n, p}.
        \intertext{\emph{Electrochemical reactions:}}
		&j\ts{k}^* =  
		j\ts{$0$,k}^* \sinh
				\left(
				  \frac{F^*\eta^*\ts{k}}{2R^*T^*} 
				\right),
		&& \kin{n, p}, \\ 
		& j\ts{$0$,k}^* = m\ts{k}^* (c\ts{s,k}^*)^{1/2} 
		(c\ts{s,k,\text{max}}^*-c\ts{s,k}^*)^{1/2}
			(c\ts{e,k}^*)^{1/2} 
		&&\kin{n, p}, \\
		&\eta^*\ts{k} = 
		\phi\ts{s,k}
			- \phi\ts{e,k}^* 
			- U\ts{k}^*(c\ts{s,k}^*\big|_{r^*=R\ts{k}^*}),
		&&\kin{n, p}.
    	\intertext{\textbf{Boundary conditions}}
		\intertext{\emph{Current:}}
		\label{eqn:dim:current:L_n} 
		& i^*\ts{e,n}\big|_{x^*=0} =
		  i^*\ts{e,p}\big|_{x^*=L^*}=0,  && \\
		\label{eqn:dim:current:L_p} 
		& \phi^*\ts{e,n}\big|_{x^*=L^*\ts{n}}
			=\phi\ts{e,s}^*\big|_{x^*=L^*\ts{n}}, \quad 
		  i^*\ts{e,n}\big|_{x^*=L^*\ts{n}}
		  	=i^*\ts{e,s}\big|_{x^*=L^*\ts{n}}
		  	=I^*, && \\ 
	  & \phi^*\ts{e,s}\big|_{x^*=L^*-L^*\ts{p}}
			=\phi^*\ts{e,p}\big|_{x^*=L^*-L^*\ts{p}}, \quad 
		  i^*\ts{e,s}\big|_{x=L^*-L^*\ts{p}}
			=i^*\ts{e,p}\big|_{x^*=L^*-L^*\ts{p}}
			=I^*. &&
		\intertext{\emph{Concentration in electrolyte:}}
		& N\ts{e,n}^*\big|_{x^*=0} = 0, \quad 
		  N\ts{e,p}^*\big|_{x^*=L^*}=0,  &&\\ 
		& c^*\ts{e,n}|_{x^*=L^*\ts{n}} = c^*\ts{e,s}|_{x^*=L^*\ts{n}}, \quad 
		  N\ts{e,n}^*\big|_{x^*=L^*\ts{n}} = N\ts{e,s}^*\big|_{x^*=L^*\ts{n}}, && \\
		& c^*\ts{e,s}|_{x^*=L^*-L^*\ts{p}} = c^*\ts{e,p}|_{x^*=L^*-L^*\ts{p}}, \quad 
		  N\ts{e,s}^*\big|_{x^*=L^*-L^*\ts{p}} = N\ts{e,p}^*\big|_{x^*=L^*-L^*\ts{p}}. && 
		\intertext{\emph{Concentration in the electrode active material:}}
		& \pdv{c\ts{s,k}^*}{r^*}\bigg|_{r^*=0} = 0, \quad 
		- D^*\ts{s,k}\pdv{c\ts{s,k}^*}{r^*}\bigg|_{r^*=R\ts{k}^*} = \frac{j\ts{k}^*}{F^*}, \quad 
		&& \kin{n, p}.
		\intertext{\textbf{Reference potential}}
		& \phi^*\ts{s,n}\big|_{x^*=0}=0.  && 
		\intertext{\textbf{Initial conditions}}
		& c\ts{s,k}^*(x^*,r^*,0) = c\ts{s,k,$0$}^*, && 
		  \kin{n, p},\\
	 	& c\ts{e,k}^*(x^*,0) = c^*\ts{e,typ}, && 
		  \kin{n, s, p}.
\end{align} 
\end{subequations}

The functional forms for $U\ts{n}^*(c\ts{s,n}^*)$, $U\ts{p}^*(c\ts{s,p}^*)$, and $D\ts{e}^*(c\ts{e}^*)$, which are the open-circuit potentials (OCPs) in the negative and positive electrodes and the lithium-ion diffusivity in the electrolyte, respectively, are taken from Newman's DUALFOIL code \cite{Dualfoil}. A full list of parameters and their values is provided in Table \ref{tab:parameters}. The values in this table are taken from Scott Moura's fastDFN code \cite{SMouraGithub}, which are in turn adapted from the parameter values used in Newman's DUALFOIL code \cite{Dualfoil}.  These functions and parameters correspond to a cell with a graphite negative electrode, a \ce{LiPF6} in EC:DMC electrolyte, and a lithium-cobalt-oxide positive electrode. In (\ref{eqn:dim:solid_ohms}), (\ref{eqn:dim:current:L_n}), and (\ref{eqn:dim:current:L_p}) the actual current density $I^*$ is distinct from the typical current density $I^*\ts{typ}$ in Table \ref{tab:parameters}. We make this distinction to easily accommodate non-constant currents within the dimensionless form of our reduced models.

\section{Asymptotic reduction of DFN model}
\subsection{Dimensionless form of DFN model}
In the following, we will use asymptotic methods to systematically reduce the DFN model to simpler forms. In order to do this, we first nondimensionalize the DFN model by making the following scalings:

\begin{equation}\label{eqn:scalings}
      \begin{aligned}  
        &\text{Global:} &&x^* = L^* x, \quad  
        t^* = \tau^*\ts{d} t, \quad 
        D^*\ts{e} = D^*\ts{e,typ}D\ts{e}, \\
        & && \kappa^*\ts{e} = \kappa\ts{e,typ}^*\kappa\ts{e} \quad
        I^* = I^*\ts{typ} I\\ 
        & && \phi^*\ts{s,n} = \frac{R^* T^*}{F^*}\phi\ts{s,n}, \quad 
           \phi^*\ts{s,p} = \left(U^*\ts{p,ref} - U^*\ts{n,ref}\right) + \frac{R^* T^*}{F^*}\phi\ts{p,k}, \\[10pt]
        &\text{For } \kin{n, p}: \ 
        &&r^*\ts{k} = R^*\ts{k} r\ts{k}, \quad 
        c\ts{s,k}^* = c^*\ts{s,k,max}c\ts{s,k}, \\
        & &&j^*\ts{k} =  \frac{I^*\ts{typ}}{a^*\ts{k}L^*}j\ts{k}, \quad j^*\ts{$0,$k} = \frac{I^*\ts{typ}}{a^*\ts{k}L^*}j\ts{$0,$k}, \quad
        m\ts{k}^* = m\ts{k,typ}^*m\ts{k}, \\
        & && \eta^*\ts{k} = \frac{R^*T^*}{F^*}\eta\ts{k}, \quad 
        U^*\ts{k} = U^*\ts{k,ref} + \frac{R^* T^*}{F^*}U\ts{k}. \quad 
         \\[10pt]
        &\text{For } \kin{n, p}: \ 
        && i^*\ts{s,k} = I^*\ts{typ} i\ts{s,k}. \\[10pt]
        &\text{For } \kin{n, s, p}: \ 
        && c^*\ts{e,k} = c^*\ts{e,typ}c\ts{e,k}, \quad 
        {N}^*\ts{e,k} = \frac{D\ts{e,typ}^*c\ts{e,typ}^*}{L^*}{N}\ts{e,k}, \\
        & && \phi^*\ts{e,k} = - U^*\ts{n,ref} + \frac{R^* T^*}{F^*}\phi\ts{e,k}, \quad 
        {i}^*\ts{e,k} = I^*\ts{typ}{i}\ts{e,k}.
    \end{aligned} 
\end{equation}
We then identify the key timescales in the model, which are presented in Table \ref{tab:timescales}. We also identify a set of dimensionless parameters, which \deleted{are provided} {\add we define} in terms of the dimensional variables \deleted{alongside}{\add together with} their physical meaning and calculated values in Table \ref{table:dimensionless_parameter_values}.

Substituting the scalings (\ref{eqn:scalings}) into the dimensional model (\ref{eqn:DFN}), we obtain the dimensionless version of the DFN model which is summarized as: 
\begin{subequations}\label{DFN}
\begin{align}
 \intertext{\textbf{Governing equations}}
    \intertext{\emph{Charge conservation}:}
    \label{eqn:nDim:electrolyte:current}
    \pdv{i\ts{e,k}}{x} &= \begin{cases}
	  j\ts{k}, \quad &\text{k} = \text{n, p},\\ 
	  0, \qquad &\text{k} = \text{s},
                        \end{cases}
                        && \kin{n, s, p}, \\
    \label{eqn:nDim:McIn} 
    i\ts{e,k} &= \epsilon\ts{k}^{\text{b}} \hat{\kappa}\ts{e}\kappa\ts{e}(c\ts{e,k}) \left( -  \pdv{\phi\ts{e,k}}{x} + 2(1-t^+)\pdv{x}\left(\log(c\ts{e,k})\right)\right), && \kin{n, s, p}, \\ 
    \label{eqn:nDim:solid:ohms}
    I-i\ts{e,k} &= - \sigma\ts{k} \pdv{\phi\ts{s,k}}{x}, &&\kin{n, p}.
    \intertext{\emph{Molar conservation}:}
    \label{eqn:nDim:electrolyte:concentration}
    \mathcal{C}\ts{e} \epsilon\ts{k} \gamma\ts{e} \pdv{c\ts{e,k}}{t} &= - \gamma\ts{e} \pdv{N\ts{e,k}}{x} + \mathcal{C}\ts{e} \pdv{i\ts{e,k}}{x}, \quad &&\kin{n, s, p}, \\ 
    \label{eqn:nDim:electrolyte:flux}
    N\ts{e,k} &= -\epsilon\ts{k}^{\text{b}}  D\ts{e}(c\ts{e,k}) \pdv{c\ts{e,k}}{x} + \frac{\mathcal{C}\ts{e} t^+}{\gamma\ts{e}} i\ts{e,k}, && \kin{n, s, p}, \\
    \label{eqn:nDim:solid:concentration}
     \mathcal{C}\ts{k}\pdv{c\ts{s,k}}{t} &= \frac{1}{r\ts{k}^2}\pdv{r\ts{k}}\left(r\ts{k}^2 \pdv{c\ts{s,k}}{r\ts{k}}\right), \quad && \kin{n, p}.
     \intertext{\emph{Electrochemical reactions}:}
      \label{eqn:nDim:BV}
     j\ts{k} &= j\ts{$0,$k} \sinh\left(\frac{ \eta\ts{k}}{2} \right), 
     &&\kin{n, p}, \\
     \label{eqn:nDim:exchange_current}
     j\ts{$0,$k} &= \frac{\gamma\ts{k}}{\mathcal{C}\ts{r,k}} c\ts{s,k}^{1/2} (1-c\ts{k})^{1/2}c\ts{e,k}^{1/2}\big|_{r\ts{k}=1}, && \kin{n, p}, \\
      \label{eqn:nDim:overpotential}
	 \eta\ts{k} &= \phi\ts{s,k} - \phi\ts{e,k} - U\ts{k}(c\ts{s,k}\big|_{r\ts{k}=1}),  && \kin{n, p}.
\end{align}
\end{subequations}
\begin{subequations}
\begin{align}
\intertext{\textbf{Boundary conditions}}
    \intertext{\emph{Current}:}
    \label{bc:nDim:no_electrolyte_cc_current}
    &i\ts{e,n}\big|_{x=0} = 0, \quad i\ts{e,p}\big|_{x=1}=0,  && \\
    \label{bc:nDim:charge_continuity_Ln}
    &\phi\ts{e,n}\big|_{x=L\ts{n}} = \phi\ts{e,s}\big|_{x=L\ts{n}}, \quad i\ts{e,n}\big|_{x=L\ts{n}} = i\ts{e,s}\big\vert_{x=L\ts{n}} = I, && \\ 
    \label{bc:nDim:charge_continuity_Lp}
    &\phi\ts{e,s}\big|_{x=1-L\ts{p}} = \phi\ts{e,p}\big|_{x=1-L\ts{p}}, \quad 
    i\ts{e,s}\big|_{x=1-L\ts{p}} = i\ts{e,p}\big|_{x=1-L\ts{p}} = I. &&
    \intertext{\emph{Concentration in electrolyte}:}
    \label{bc:nDim:no_electrolyte_cc_flux}
    &N\ts{e,n}\big|_{x=0} = 0, \quad N\ts{e,p}\big|_{x=1}=0,  &&\\ 
    \label{bc:nDim:concentration_continuity_Ln}
    &c\ts{e,n}\big|_{x=L\ts{n}} = c\ts{e,s}|_{x=L\ts{n}}, \quad N\ts{e,n}\big|_{x=L\ts{n}}=N\ts{e,s}\big|_{x=L\ts{n}}, && \\
    \label{bc:nDim:concentration_continuity_Lp}
    &c\ts{e,s}|_{x=1-L\ts{p}}=c\ts{e,p}|_{x=1-L\ts{p}}, \quad N\ts{e,s}\big|_{x=1-L\ts{p}}=N\ts{e,p}\big|_{x=1-L\ts{p}}. && 
    \intertext{\emph{Concentration in the electrode active material}:}
    \label{bc:nDim:particle}
    & \pdv{c\ts{s,k}}{r\ts{k}}\bigg|_{r\ts{k}=0} = 0, \quad -\frac{a\ts{k}\gamma\ts{k}}{\mathcal{C}\ts{k}}\pdv{c\ts{s,k}}{r\ts{s,k}}\bigg|_{r\ts{k}=1} = j\ts{k}, \quad \kin{n, p}. &&
    \intertext{\emph{Reference potenital}:}
    & \phi\ts{s,n}\big|_{x=0} = 0. && \label{bc:nDim:reference_potential} 
    \intertext{\textbf{Initial conditions}}
     \label{ic:nDim:solid} 
	&c\ts{s,k}(x,r,0) = c\ts{s,k,0}, && \kin{n, p},\\
	\label{ic:nDim:electrolyte}
	& c\ts{e,k}(x,0) = 1, && \kin{n, s, p}.
\end{align}
\end{subequations}

Before proceeding with model reduction, we note some helpful relations. The terminal voltage is given by
\begin{equation}
  V = \phi\ts{s,p}\big|_{x=1} - \phi\ts{s,n}\big|_{x=0}.
\end{equation}
We can re-write this in a more useful form by considering a \deleted{general}{\add particular} path {\add that the} current follows through the cell. We consider the current entering through the negative current collector at $x=0$ and travelling through the solid phase of the negative electrode to a point $x\ts{n}\in[0,L\ts{n}]$. At this point, an electrochemical reaction occurs so that the current is transferred into the electrolyte. The current then travels through the electrolyte until it reaches a point $x\ts{p}\in[1-L\ts{p}, 1]$ where another electrochemical reaction occurs transferring the current into the solid phase of the positive electrode. Finally, the current travels through the positive electrode until it reaches the positive current collector. The terminal voltage can be written in terms of the potential drops associated with each section of this path as
\begin{equation}\label{eqn:expanded-voltage}
  \begin{aligned}
  V =& \phantom{+}  \phi\ts{s,n}\big|_{x=x\ts{n}} - \phi\ts{s,n}\big|_{x=0} \quad &&\text{(Negative electrode)} \\
      &+ \phi\ts{e,n}\big|_{x=x\ts{n}} - \phi\ts{s,n}\big|_{x=x\ts{n}} \quad && \text{(Negative electrochemical reaction)} \\
      &+ \phi\ts{e,p}\big|_{x=x\ts{p}} - \phi\ts{e,n}\big|_{x=x\ts{n}} \quad && \text{(Electrolyte)} \\
      &+ \phi\ts{s,p}\big|_{x=x\ts{p}} - \phi\ts{e,p}\big|_{x=x\ts{p}} \quad && \text{(Positive electrochemical reaction)} \\
      &+ \phi\ts{s,p}\big|_{x=1} - \phi\ts{s,p}\big|_{x=x\ts{p}} \quad && \text{(Positive electrode).} \\
  \end{aligned}
\end{equation}
We define the pointwise OCV to be
\begin{equation}
    U\ts{eq}\big|_{x\ts{n},x\ts{p}} = U\ts{p}(c\ts{s,p}\big|_{r=1})\big|_{x=x\ts{p}} - U\ts{n}(c\ts{s,n}\big|_{r=1})\big|_{x=x\ts{n}},
\end{equation}
and the pointwise solid-phase Ohmic losses to be
\begin{equation}
  \Delta \Phi\ts{Solid}\big|_{x\ts{n},x\ts{p}} = 
        \left(\phi\ts{s,p}\big|_{x=1} - \phi\ts{s,p}\big|_{x=x\ts{p}} \right) + \left(\phi\ts{s,n}\big|_{x=x\ts{n}} - \phi\ts{s,n}\big|_{x=0}\right).
\end{equation}
Then, using the definition of the reaction overpotential given in (\ref{eqn:nDim:overpotential}), {\add equation} (\ref{eqn:expanded-voltage}) becomes
\begin{equation}\label{eqn:pointwise-voltage}
    V = U\ts{eq}\big|_{x\ts{n},x\ts{p}} 
        + \eta\ts{p}\big|_{x=x\ts{p}} - \eta\ts{n}\big|_{x=x\ts{n}} 
        + \phi\ts{e,p}\big|_{x=x\ts{p}} - \phi\ts{e,n}\big|_{x=x\ts{n}} 
        + \Delta \Phi\ts{Solid}\big|_{x\ts{n},x\ts{p}} .
\end{equation}
Equation (\ref{eqn:pointwise-voltage}) provides the voltage in terms the internal pointwise OCV, Ohmic losses, and overpotentials for a particular current path through the cell. However, we can also express the voltage in terms of the average (across all paths) OCV, Ohmic losses, and overpotentials by averaging (\ref{eqn:pointwise-voltage}) over each electrode. This is done by integrating (\ref{eqn:pointwise-voltage}) with respect to $x\ts{n}\in[0, L\ts{n}]$ and dividing by the negative electrode thickness, $L\ts{n}$, and then integrating with respect to $x\ts{p}\in[1-L\ts{p}, 1]$ and dividing by the positive electrode thickness, $L\ts{p}$. The result is that we can express the terminal voltage as
\begin{equation}\label{eqn:electrode-averaged-voltage}
  V = \mybar{U}\ts{eq}
        + \mybar{\eta}\ts{p} - \mybar{\eta}\ts{n}
        + \mybar{\phi}\ts{e,p} - \mybar{\phi}\ts{e,n}
        + \mybar{\Delta \Phi}\ts{Solid}.
\end{equation}
where we use an overbar to represent the operation
\begin{equation}
  \mybar{f} := \frac{1}{L\ts{p}}\int_{1}^{1-L\ts{p}} 
                  \left(\frac{1}{L\ts{n}}\int_0^{L\ts{n}} f \, \text{d}x\ts{n} \right) \, \text{d}x\ts{p}.
\end{equation}

It is this electrode-averaged form of the voltage expression that will play a central role in our extension of the SPM. Without this electrode-averaged form, the extended model would need to be more complex than the model we present in order to obtain the same degree of accuracy.

Another useful property can be found by integrating (\ref{eqn:nDim:electrolyte:concentration}) with respect to $x$ over $\Omega\ts{k}$ for each $\kin{n, s, p}$, applying the boundary conditions (\ref{bc:nDim:no_electrolyte_cc_current}), (\ref{bc:nDim:charge_continuity_Ln}), (\ref{bc:nDim:charge_continuity_Lp}), (\ref{bc:nDim:no_electrolyte_cc_flux}), (\ref{bc:nDim:concentration_continuity_Ln}), (\ref{bc:nDim:concentration_continuity_Lp}), and applying the initial condition (\ref{ic:nDim:electrolyte}),  to obtain
\begin{equation}\label{eqn:nDim:lithiumCondition} 
    \int_{0}^{L\ts{n}}c\ts{e,n}\,dx + \int_{L\ts{n}}^{1-L\ts{p}} c\ts{e,s} \, dx + \int_{1-L\ts{p}}^{1} c\ts{e,p}\,dx = 1.
\end{equation}
Equation (\ref{eqn:nDim:lithiumCondition}) is a statement of the conservation of the total number of lithium ions in the electrolyte.

\subsection{The limit $\mathcal{C}\ts{e} \rightarrow0$}\label{sec:asymptotic-reduction}
We consider the limit of high electrical conductivity in the electrodes and electrolyte (such that the typical potential drop in each material is small relative to the thermal voltage) and the timescale for the migration of lithium ions in the electrolyte is small relative to the typical timescale of a discharge. This corresponds to the limit $\mathcal{C}\ts{e}\rightarrow0$, where $\mathcal{C}\ts{e}$ is the ratio of the typical timescale for lithium-ion migration to the typical discharge timescale, $\sigma\ts{k}\rightarrow\infty$, where $\sigma\ts{k}$ is ratio of the thermal voltage to the typical Ohmic drop in the solid, and $\hat{\kappa}\ts{e}\rightarrow\infty$, where $\hat{\kappa}\ts{e}$ is the ratio of the thermal voltage to the typical Ohmic drop in the electrolyte. We take the distinguished limit in which $\sigma\ts{k}\mathcal{C}\ts{e}$ and $\hat{\kappa}\ts{e}\mathcal{C}\ts{e}$ both tend to a constant as $\mathcal{C}\ts{e}\rightarrow 0$, $\sigma\ts{k}\rightarrow\infty$, and $\hat{\kappa}\ts{e}\rightarrow\infty$ by setting 
    \begin{align*}
    &\sigma\ts{k} = \frac{\sigma\ts{k}'}{\mathcal{C}\ts{e}}, \quad \sigma\ts{k}' = \mathcal{O}(1), \quad \kin{n, p}, \\ 
    &\hat{\kappa}\ts{e} = \frac{\hat{\kappa}\ts{e}'}{\mathcal{C}\ts{e}}, \quad \hat{\kappa}\ts{e}' = \mathcal{O}(1).
    \end{align*}
We then expand all variables in powers of $\mathcal{C}\ts{e}$ in the form
\begin{align*} 
  c\ts{s,k} \sim c\ts{s,k}^0 + \mathcal{C}\ts{e} c\ts{s,k}^1 + \mathcal{C}\ts{e}^2 c\ts{s,k}^2 + \dots ,
\end{align*}
etc.

\subsubsection{Leading-order model}
At leading order in $\mathcal{C}\ts{e}$, (\ref{eqn:nDim:electrolyte:concentration}) and (\ref{eqn:nDim:electrolyte:flux}) are
\begin{equation} \label{eqn:leading-order-lithium-ion-fluxes}
  \pdv{N\ts{e,k}^0}{x} = 0, \quad N\ts{e,k}^0 = -\epsilon\ts{k}^bD\ts{e}(c\ts{e,k}^0) \pdv{c\ts{e,k}^0}{x}, \quad \kin{n, s, p}.
\end{equation}
Upon application of the leading-order boundary conditions (\ref{bc:nDim:no_electrolyte_cc_flux}), (\ref{bc:nDim:concentration_continuity_Ln}), (\ref{bc:nDim:concentration_continuity_Lp}), the leading-order initial condition (\ref{ic:nDim:electrolyte}), and the leading-order component of the condition (\ref{eqn:nDim:lithiumCondition}), equation (\ref{eqn:leading-order-lithium-ion-fluxes}) yields
\begin{equation}
  N\ts{e,k}^0 = 0, \quad c\ts{e,k}^0 = 1, \quad \kin{n, s, p}.
\end{equation}
Thus there is no depletion of the electrolyte at leading order. Equations (\ref{eqn:nDim:McIn}) and (\ref{eqn:nDim:solid:ohms}), at leading order in $\mathcal{C}\ts{e}$, are then 
\begin{equation}\label{eqn:lo:phi_const}
	\pdv{\phi\ts{e,k}^0}{x} = 0, \quad \pdv{\phi\ts{s,k}^0}{x} = 0.
\end{equation}
Since $c\ts{e,k}^0$, $\phi\ts{e,k}^0$, and $\phi\ts{s,k}^0$ are all independent of $x$ and $c\ts{s,k}^0$ is initially independent of $x$, it follows from the leading-order components of (\ref{eqn:nDim:BV}), (\ref{eqn:nDim:exchange_current}), and (\ref{eqn:nDim:overpotential}), that $j\ts{k}^0$, $j\ts{$0$,k}^0$, and $\eta\ts{k}^0$ do not depend on $x$. Therefore, we integrate (\ref{eqn:nDim:electrolyte:current}) with respect to $x$ over $\Omega\ts{k}$ for $\text{k}=\text{n, p}$ and apply (\ref{bc:nDim:no_electrolyte_cc_current}), (\ref{bc:nDim:charge_continuity_Ln}), (\ref{bc:nDim:charge_continuity_Lp}), to obtain
\begin{align}
    \label{eqn:lo:current} 
	&i\ts{e,n}^0 = \frac{xI}{L\ts{n}}, \quad i\ts{e,s}^0=I, \quad i\ts{e,p}^0 = \frac{(1-x)I}{L\ts{p}}, \\ 
    \label{eqn:lo:G0}
	&j\ts{n}^0 = \frac{I}{L\ts{n}}, \quad j\ts{p}^0 = -\frac{I}{L\ts{p}}.
\end{align}
From the leading-order component of (\ref{eqn:nDim:BV}), we then have 
\begin{align}\label{eqn:lo:eta}
  \eta\ts{n}^0 = 2\sinh^{-1}\left(\frac{I}{j\ts{$0$,n}^0L\ts{n}} \right), \quad \eta\ts{p}^0 =- 2\sinh^{-1}\left(\frac{I}{j\ts{0,p}^0L\ts{p}} \right).
\end{align}
Employing (\ref{eqn:lo:phi_const}) in conjunction with the leading-order interior boundary conditions (\ref{bc:nDim:charge_continuity_Ln}) and (\ref{bc:nDim:charge_continuity_Lp}), we obtain 
\begin{equation}\label{eqn:lo:concentration:overportential}
    \phi\ts{e,p}^0 - \phi\ts{e,n}^0 = 0,
\end{equation}
so {\add that} to leading order there is no potential drop in the electrolyte (for both the pointwise and electrode-averaged cases). Further, (\ref{eqn:lo:phi_const}) \deleted{we also have} {\add implies} that there are no solid-phase Ohmic losses (for both the pointwise and electrode-averaged cases) at leading order.

We are now in position to summarize the leading-order model. The leading-order description of lithium in the electrode particles is given by taking the leading-order component of (\ref{eqn:nDim:solid:concentration}) and  inserting (\ref{eqn:lo:G0}) into the leading-order component of (\ref{bc:nDim:particle}) to obtain 
\begin{subequations}\label{eqn:SPM}
	\begin{align} \label{eqn:lo:solidConcentration} 
     \mathcal{C}\ts{k}\pdv{c\ts{s,k}^0}{t} &= \frac{1}{r\ts{k}^2}\pdv{r\ts{k}}\left(r\ts{k}^2 \pdv{c\ts{s,k}^0}{r\ts{k}}\right), && \kin{n, p}, \\
     \pdv{c\ts{s,k}^0}{r\ts{k}}\bigg|_{r\ts{k}=0} &= 0, \quad -\frac{a\ts{k}\gamma\ts{k}}{\mathcal{C}\ts{k}}\pdv{c\ts{s,k}^0}{r\ts{k}}\bigg|_{r\ts{k}=1} = 
     \begin{cases}
     \frac{I}{L\ts{n}}, \quad & \text{k}=\text{n}, \\ 
     -\frac{I}{L\ts{p}}, \quad &\text{k}=\text{p}, 
     \end{cases} \label{eqn:lo:solidConcentration:BC} 
     && \kin{n, p}, \\
     c\ts{s,k}^0(r\ts{k},0) &= c\ts{s,k,0}, && \kin{n, p}.
    \end{align} 
Since $c\ts{s,k}^0$ is independent of $x$, the leading-order electrode-averaged OCV is simply
\begin{equation}
  \mybar{U}\ts{eq}^0 = U\ts{p}(c\ts{s,p}^0\big|_{r\ts{p}=1})-U\ts{n}(c\ts{s,n}^0\big|_{r\ts{n}=1}).
\end{equation}
Additionally, the leading-order electrode-averaged reaction overpotentials are just $\mybar{\eta}^0\ts{k} = \eta\ts{k}^0$. Therefore, the leading-order voltage is given by
\begin{equation}
    V^0 = 
    \underbrace{\Bigg. U\ts{p}(c\ts{s,p}^0)\big|_{r\ts{p}=1}-U\ts{n}(c\ts{s,n}^0)\big|_{r\ts{n}=1}}_{\text{\normalfont OCV}} 
    \underbrace{\Bigg. -2\sinh^{-1}\left(\frac{I}{j\ts{0,p}^0L\ts{p}} \right)  -2\sinh^{-1}\left(\frac{I}{j\ts{0,n}^0L\ts{n}} \right)}_{\text{Reaction overpotentials}},
\end{equation}
where the leading-order component of the exchange-current density, as given by (\ref{eqn:nDim:exchange_current}), is 
\begin{equation}\label{eqn:lo:exchange_current}
    j\ts{$0$,k}^0=\frac{\gamma\ts{k}}{\mathcal{C}\ts{r,k}} (c^0\ts{s,k})^{1/2} (1-c\ts{s,k}^0)^{1/2}.
\end{equation}
\end{subequations} 

We identify (\ref{eqn:SPM}) as the dimensionless form of the SPM \cite{Perez2016, Bizeray2017}. The name refers to the requirement to only solve a diffusion equation in one particle in each electrode, rather than solving a diffusion equation a particle at every macroscopic point as in the DFN model. This model should not be interpreted as replacing the many particles in an electrode by a single particle. Instead, in this limit all the particles in an electrode behave in exactly the same way and it is therefore sufficient to solve for just one representative particle.

\subsubsection{First-order correction}
We now proceed to calculate the $\mathcal{O}(\mathcal{C}\ts{e})$ corrections terms, which will extend the range of applicability of the reduced-order model to higher C-rates. At $\mathcal{O}(\mathcal{C}\ts{e})$, (\ref{eqn:nDim:electrolyte:concentration}) and (\ref{eqn:nDim:electrolyte:flux}) become
\begin{align}\label{eqn:od:flux_deriv}
  &\pdv{N\ts{e,k}^1}{x} = \frac{1}{\gamma\ts{e}}\pdv{i\ts{e,k}^0}{x}, && \kin{n, p}, \\
    \label{eqn:od:flux}
	&N\ts{e,k}^1 = -\epsilon\ts{k}^b D\ts{e}(c^0\ts{e,k}) \pdv{c^1\ts{e,k}}{x} + \frac{t^+}{\gamma\ts{e}}i\ts{e,k}^0, && \kin{n, p}.
\end{align}
Integrating (\ref{eqn:od:flux_deriv}) with respect to $x$ over $\Omega\ts{k}$ for each $\kin{n, s, p}$ and applying the $\mathcal{O}(\mathcal{C}\ts{e})$ components of the boundary conditions (\ref{bc:nDim:no_electrolyte_cc_flux}), (\ref{bc:nDim:concentration_continuity_Ln}), and (\ref{bc:nDim:concentration_continuity_Lp}), we obtain
\begin{equation}\label{eqn:od:fluxes:sol} 
  N\ts{e,n} = \frac{I x}{\gamma\ts{e} L\ts{n}}, \quad N\ts{e,s} = \frac{I}{\gamma\ts{e}}, \quad N\ts{e,p} = \frac{I (1-x)}{\gamma\ts{e} L\ts{p}}. 
\end{equation}
We substitute (\ref{eqn:od:fluxes:sol}) into (\ref{eqn:od:flux}) and then integrate with respect to $x$ over $\Omega\ts{k}$ for each $\kin{n, s, p}$ using the $\mathcal{O}(\mathcal{C}\ts{e})$ components of the continuity boundary conditions (\ref{bc:nDim:concentration_continuity_Ln}), (\ref{bc:nDim:concentration_continuity_Lp}), and lithium-ion conservation condition (\ref{eqn:nDim:lithiumCondition}) to determine the constants of integration. From this, we get
\begin{subequations}
\begin{align}
	&c\ts{e,n}^1 = \frac{(1-t^+) I}{\gamma\ts{e} 6 D\ts{e}(1)}
        \left( 2\left(\frac{L\ts{p}^2}{\epsilon\ts{p}^b} - \frac{L\ts{n}^2}{\epsilon\ts{n}^b}\right)
         +\frac{3L_{s}}{\epsilon\ts{s}^b}(1+L\ts{p}-L\ts{n}) + \frac{3}{\epsilon\ts{n}^b L\ts{n}}(L\ts{n}^2-x^2)
         \right), \\ 
	&c\ts{e,p}^1 = \frac{(1-t^+) I}{ 6\gamma\ts{e} D\ts{e}(1)}
        \left( 2\left(\frac{L\ts{p}^2}{\epsilon\ts{p}^b} - \frac{L\ts{n}^2}{\epsilon\ts{n}^b}\right)
         +\frac{3}{\epsilon\ts{s}^b}(L\ts{n}^2 - L\ts{p}^2 + 1-2x)
         \right), \\
	&c\ts{e,p}^1 = \frac{(1-t^+) I}{ 6 \gamma\ts{e} D\ts{e}(1)}
        \left( 2\left(\frac{L\ts{p}^2}{\epsilon\ts{p}^b} - \frac{L\ts{n}^2}{\epsilon\ts{n}^b}\right)
         +\frac{3L_{s}}{\epsilon\ts{s}^b}(L\ts{p} - L\ts{n} - 1) + \frac{3}{L\ts{p} \epsilon\ts{p}^b}( (x-1)^2 - L\ts{p}^2 )
         \right).
\end{align}
\end{subequations}

At $\mathcal{O}(\mathcal{C}\ts{e})$, (\ref{eqn:nDim:McIn}) becomes
\begin{gather}\label{eqn:od:current_to_integrate}
  i\ts{e,k}^0 = \epsilon\ts{k}^b \hat{\kappa}\ts{e}' \kappa\ts{e}(1) \left( -\pdv{\phi\ts{e,k}^1}{x} + 2(1-t^+)\pdv{c\ts{e,k}^1}{x} \right).
\end{gather} 
We substitute (\ref{eqn:lo:current}) into (\ref{eqn:od:current_to_integrate}), integrate with respect to $x$ over $\Omega\ts{k}$ for each $\kin{n, s, p}$, and determine two of the three constants of integration by applying the $\mathcal{O}(\mathcal{C}\ts{e})$ components of the interior boundary conditions (\ref{bc:nDim:charge_continuity_Ln}) and (\ref{bc:nDim:concentration_continuity_Lp}) to obtain
\begin{subequations}\label{eqn:second-order-phi} 
    \begin{align}
		&\phi\ts{e,n}^1 = \tilde{\phi}\ts{e} + 2(1-t^+)c\ts{e,n}^1
		- \frac{ I}{\hat{\kappa}\ts{e}' \kappa\ts{e}(1)}\left(\frac{x^2-L\ts{n}^2}{2 \epsilon\ts{n}^b L\ts{n}} + \frac{L\ts{n}}{\epsilon\ts{s}^b} \right), \\
		&\phi\ts{e,s}^1 = \tilde{\phi}\ts{e}  + 2(1-t^+)c\ts{e,s}^1
		- \frac{ I x}{\hat{\kappa}\ts{e}' \kappa\ts{e}(1)\epsilon\ts{s}^b},\\
		&\phi\ts{e,n}^1 = \tilde{\phi}\ts{e} + 2(1-t^+)c\ts{e,p}^1
		- \frac{I}{\hat{\kappa}\ts{e}' \kappa\ts{e}(1)}\left(\frac{x(2-x)+L\ts{p}^2-1}{2\epsilon\ts{p}^b L\ts{p}} + \frac{1-L\ts{p}}{\epsilon\ts{s}^b} \right),
    \end{align}
\end{subequations} 
where $\tilde{\phi}\ts{e}$ is a constant, the form of which is provided in Appendix \ref{app:electrolyte-constants}. 

At $\mathcal{O}(\mathcal{C}\ts{e})$, (\ref{eqn:nDim:solid:ohms}) is
\begin{equation}
  I - i\ts{e,k}^0 = -\sigma'\ts{k} \pdv{\phi\ts{s,k}^1}{x}, \quad \kin{n, p},
\end{equation}
which upon using (\ref{eqn:lo:current}) and integrating with respect to $x$, gives 
\begin{subequations}\label{eqn:od-solid-potentials}
\begin{align}
  &\phi\ts{s,n}^1 = \phi\ts{s,n}\big|_{x=0} + \frac{Ix}{2\sigma'\ts{n} L\ts{n}}\left(2L\ts{n} - x\right),\\ 
	&\phi\ts{s,p}^1 = \phi\ts{s,p}\big|_{x=1} + \frac{I (x-1) (1-2L\ts{p}-x)}{2\sigma'\ts{p} L\ts{p}}.
\end{align}
\end{subequations} 

At $\mathcal{O}(\mathcal{C}\ts{e})$, (\ref{eqn:nDim:electrolyte:current}), (\ref{bc:nDim:no_electrolyte_cc_current}), (\ref{bc:nDim:charge_continuity_Ln}), and (\ref{bc:nDim:charge_continuity_Lp}) give
\begin{subequations} 
\begin{align}
  &\pdv{i\ts{e,k}^1}{x} = j\ts{k}^1, &&\kin{n, p}, \label{eqn:od:elect:curr}\\ 
	&i\ts{e,n}^1\big|_{x=0,L\ts{n}} = i\ts{e,p}^1\big|_{1-L\ts{p},1} = 0. \label{eqn:od:elect:curr:bc}
\end{align} 
\end{subequations}
Here we approach a key step in our derivation. Integrating (\ref{eqn:od:elect:curr}) with respect to $x$ over $\Omega\ts{k}$ for $\kin{n, p}$ and applying (\ref{eqn:od:elect:curr:bc}) gives the conditions
\begin{equation}\label{eqn:od:conditions} 
  \int_0^{L\ts{n}} j\ts{n}^1\, \text{d}x = 0, \ \text{and} \ \int_{1-l\ts{p}}^1 j\ts{p}^1\, \text{d}x = 0. 
\end{equation}
Thus the electrode-averaged corrections to the reaction currents $j\ts{n}$ and $j\ts{p}$ are zero. This means that after averaging the $\mathcal{O}(\mathcal{C}\ts{e})$ components of (\ref{eqn:nDim:solid:concentration}) and  (\ref{bc:nDim:particle}) in $x$ over $\Omega\ts{k}$ for each $\kin{n, p}$ and using (\ref{eqn:od:conditions}), we get
\begin{subequations}\label{eqn:od:electrode_av_con}
\begin{align}
  &\mathcal{C}\ts{k}\pdv{\mybar{c}\ts{s,k}^1}{t} = \frac{1}{r\ts{k}^2}\pdv{r\ts{k}}\left(r\ts{k}^2 \pdv{\mybar{c}\ts{s,k}^1}{r\ts{k}}\right),  && \kin{n, p},\\
	 &\pdv{\mybar{c}\ts{s,k}^1}{r\ts{k}}\bigg|_{r\ts{k}=0} = 
	 \pdv{\mybar{c}\ts{s,k}^1}{r\ts{k}}\bigg|_{r\ts{k}=1} = 0, &&\kin{n, p},\\ 
	 &\mybar{c}\ts{s,k}^1(r\ts{k}, 0) = 0, && \kin{n, p}.
\end{align} 
\end{subequations} 
Crucially, there is no average flux on the surface $r\ts{k}=1$ in (\ref{eqn:od:electrode_av_con}). The solution to (\ref{eqn:od:electrode_av_con}) is therefore simply
\begin{equation}\label{eqn:od:av_con_sol} 
  \mybar{c}\ts{s,n}^1 = 0, \quad \mybar{c}\ts{s,p}^1 = 0.
\end{equation}
Therefore, the $\mathcal{O}(\mathcal{C}\ts{e})$ component of the electrode-averaged OCV is
\begin{equation}\label{eqn:ueq-bar-1}
  \mybar{U}\ts{eq}^1 = U\ts{p}'(c\ts{s,p}^0)\mybar{c}\ts{s,p}^1 - 
                      U\ts{n}'(c\ts{s,n}^0)\mybar{c}\ts{s,n}^1
                  = 0.
\end{equation}
The $\mathcal{O}(\mathcal{C}\ts{e})$ components of (\ref{eqn:nDim:BV}), (\ref{eqn:nDim:overpotential}), and (\ref{eqn:nDim:exchange_current}) are 
\begin{subequations}\label{eqn:od:all_BV}
    \begin{align} \label{eqn:od:BV}
    	&j\ts{k}^1 = j\ts{$0$,k}^1 \sinh\left( \frac{\eta\ts{k}^0}{2}  \right) + \frac{j\ts{$0$,k}^0  \eta\ts{k}^1}{2}\cosh(\frac{\eta\ts{k}^0}{2}), \\
    	&\eta\ts{k}^1 = \phi^1\ts{s,k} - \phi^1\ts{e,k}  - U\ts{k}'(c^0\ts{s,k})c\ts{s,k}^1\big|_{r\ts{k}=1},\\
        & j\ts{$0$,k}^1 = \frac{j\ts{$0$,k}^0}{2}\left(\frac{c\ts{s,k}^1}{c\ts{s,k}^0} - \frac{c\ts{s,k}^1}{1-c\ts{s,k}^0} + c\ts{e,k}^1 \right)\bigg|_{r\ts{k}=1},
        \end{align} 
\end{subequations}
where $\eta\ts{k}^0$ and $j\ts{$0$,k}^0$ are given by (\ref{eqn:lo:eta}) and (\ref{eqn:lo:exchange_current}), respectively. Averaging (\ref{eqn:od:BV}) over electrode $\Omega\ts{k}$, gives
\begin{equation}\label{eqn:eta-bar-1}
  \begin{aligned}
    \mybar{\eta}\ts{k}^1 = -\mybar{c}\ts{e,k}^1 \tanh\left(\frac{\eta\ts{k}^0}{2}\right)= \frac{\mybar{c}\ts{e,k}^1I}{\sqrt{(j_{0,k}^0L_k)^2 + I^2}}.
  \end{aligned}
\end{equation}
We can also easily obtain $\mybar{\phi}\ts{e,k}^1$ and $\mybar{\phi}\ts{s,k}^1$ by electrode-averaging (\ref{eqn:second-order-phi}) and (\ref{eqn:od-solid-potentials}). The expressions for these can be found in Appendix \ref{app:electrode_averaged}. 

We now have all the components necessary to obtain the first-order correction to the voltage. Inserting (\ref{eqn:ueq-bar-1}), (\ref{eqn:eta-bar-1}), and the electrode-averaged potential expressions in Appendix \ref{app:electrode_averaged} into (\ref{eqn:electrode-averaged-voltage}), we obtain
\begin{equation}\label{eqn:voltage-correction}
  \begin{aligned}
	V^1 = & \left(\frac{\mybar{c}\ts{e,p}I}{\sqrt{\big. \smash[b]{(j\ts{0,p}^0L\ts{p})^2 + I^2}}} 
          + \frac{\mybar{c}\ts{e,n}I}{\sqrt{\big. (j\ts{0,n}^0L\ts{n})^2 + I^2}} \right)
           + 2(1-t^+)\left(\mybar{c}\ts{e,p}^1 - \mybar{c}\ts{e,n}^1\right) \\
	      & -\frac{I}{\hat{\kappa}\ts{e}' \kappa\ts{e}(1)}\left(\frac{L\ts{n}}{3\epsilon\ts{n}^b} + \frac{L\ts{s}}{\epsilon\ts{s}^b} + \frac{L\ts{p}}{3\epsilon\ts{p}^b} \right)
         - \frac{I}{3}\left(\frac{L\ts{p}}{\tilde{\sigma}\ts{p}} 
        + \frac{L\ts{n}}{\tilde{\sigma}\ts{n}}\right).
  \end{aligned}
\end{equation}
This is a purely algebraic expression and is therefore obtained without the numerical solution of any additional PDEs. As a result, no additional states need to be stored at each time step, which is beneficial for microcontrollers where RAM is limited. Further, the additional computational cost compound to the SPM (\ref{eqn:SPM}) is evaluating this algebraic expression and is therefore negligible. The first term of (\ref{eqn:voltage-correction}) corresponds to corrections to the reaction overpotentials due to concentration variations in the electrolyte, the second to the concentration overpotential, the third to electrolyte Ohmic losses, and the fourth to solid-phase Ohmic losses. 

\subsection{Combined voltage expression}
To write an expression for the combined leading- and first-order voltage, $V=V^0 + \mathcal{C}\ts{e} V^1$, 
we define the electrode-averaged exchange-current densities to be 
\begin{subequations}
\begin{align}
  &\mybar{j}\ts{0,n} =  \frac{1}{L\ts{n}}\int_0^{L\ts{n}} \frac{\gamma\ts{n}}{\mathcal{C}\ts{r,n}} (c\ts{s,n}^0)^{1/2}(1-c\ts{s,n}^0)^{1/2} (1+\mathcal{C}\ts{e} c\ts{e,n}^1 )^{1/2} \, \text{d}x, \\
	&\mybar{j}\ts{0,p} =  \frac{1}{L\ts{p}}\int_{1-L\ts{p}}^{1} \frac{\gamma\ts{p}}{\mathcal{C}\ts{r,p}}  (c\ts{s,p}^0)^{1/2}(1-c\ts{s,p}^0)^{1/2} (1+\mathcal{C}\ts{e} c\ts{e,p}^1 )^{1/2} \, \text{d}x,
\end{align}
\end{subequations}
and use the fact that 
\begin{align*} 
  &-2\sinh^{-1}\left(\frac{I}{\mybar{j}\ts{0,k} L\ts{k}}\right) = -2\sinh^{-1}\left(\frac{I}{j\ts{0,k}^0L\ts{k}} \right)
	+\mathcal{C}\ts{e} \frac{\mybar{c}\ts{e,k}^1 I}{\lambda\sqrt{(j\ts{$0$,k}^0L\ts{k})^2+I^2}} 
    + \mathcal{O}\left(\mathcal{C}\ts{e}^2\right).
\end{align*} 
The combined voltage accurate to $\mathcal{O}\left(\mathcal{C}\ts{e}^2\right)$ is then
\begin{equation}
  V = \mybar{U}\ts{eq} + \mybar{\eta}\ts{r}
   + \mybar{\eta}\ts{c} 
   + \mybar{\Delta\Phi}_{\text{Elec}}
   + \mybar{\Delta\Phi}_{\text{Solid}},
\end{equation}
with the combined leading- and first-order electrode-averaged components of the voltage expression given by
\begin{subequations}\label{eqn:ohms_and_overpotential}
\begin{align}
  & \mybar{U}\ts{eq} = U\ts{p}(c\ts{s,p}^0\big|_{r\ts{p}=1} )- U\ts{n}(c\ts{s,n}^0\big|_{r\ts{n}=1}),\\ 
  & \mybar{\eta}\ts{r} = -2\sinh^{-1}\left(\frac{I}{\mybar{j}\ts{p} L\ts{n}}\right)
	 -2\sinh^{-1}\left(\frac{I}{\mybar{j}\ts{n} L\ts{n}}\right), \\
	& \mybar{\eta}\ts{c} =  2 \mathcal{C}\ts{e} (1-t^+)\left(\mybar{c}\ts{e,p}^1 - \mybar{c}\ts{e,n}^1\right), \\
	&\mybar{\Delta \Phi}_{\text{Elec}}= -\frac{I}{\hat{\kappa}\ts{e} \kappa\ts{e}(1)}\left(\frac{L\ts{n}}{3\epsilon\ts{n}^b} + \frac{L\ts{s}}{\epsilon\ts{s}^b} + \frac{L\ts{p}}{3\epsilon\ts{p}^b} \right), \\
    &\mybar{\Delta \Phi}_{\text{Solid}} = -\mathcal{C}\ts{e}(\mybar{\varphi}\ts{s,p}^1 - \mybar{\phi}\ts{s,n}^1).
\end{align}
\end{subequations}
We refer to this model that includes the first-order correction to the terminal voltage as the SPMe(S). That is, the single particle model with electrolyte with the (S) referring to the fact that this formulation considers the electrolyte to be in steady state. 

Before proceeding, we would like to draw further attention to what we view as the key step in the derivation of the SPMe(S), namely the electrode-averaging step. Electrode averaging is essential as it provides us with a well-defined problem: given the electrode-averaged current, find the electrode-averaged potential differences. If instead we try to evaluate the pointwise voltage expression, we must determine both the current (since we only know the electrode-averaged current) and potential difference at a particular point. To get around this issue, it is typical for ad-hoc models in the literature, which use a pointwise voltage expression, to implicitly assume that the current at a particular point is equal to the electrode-averaged current. This is not in general true, since it implies that the concentrations in every particle of the DFN model would be the same across all operating conditions. Of course, this is not a bad first assumption to make in the limit we have been considering, and it is indeed valid for very low currents; this is the reason the SPM is a reasonable approximation. However, we have shown systematically in our asymptotic expansion that this assumption is only true at leading order and not at first order. Electrode averaging removes the requirement for this assumption, and ensures our expressions are also valid at first order, so we gain additional accuracy over the ad-hoc models for negligible additional computational cost.

\section{Canonical SPMe} 
The SPMe(S) holds for cases where the electrolyte can be taken to be in quasi-steady state (e.g. when the current varies over a longer timescale than the electrolyte diffusion timescale). However, for many applications transient effects in the electrolyte are important, particularly after a step change in current. These transient effects occur on the timescale of migration of lithium ions in the electrolyte. To include these effects it is therefore natural to scale time with the timescale for migration of lithium ions, $t^*=\tau\ts{e}\tilde{t}$. On this short migration-timescale, at leading and first order the concentrations in the electrode particles remain constant and the terms describing exchange of lithium within the electrolyte equation are negligible. We do not present the corresponding systematic asymptotic analysis for this migration-timescale problem here but instead simply state the composite model which produces the correct result on both the diffusion and discharge timescales. The main difference between this model and the SPMe(S) is that we must now solve a PDE to obtain the first-order correction for the electrolyte concentration, $c\ts{e,k}^1$. We shall take this model to be the canonical SPMe and therefore simply refer to it as the SPMe. The SPMe is summarized as:

\begin{subequations}\label{eqn:SPMe}
    \begin{align}
    \intertext{\textbf{Governing equations}}
	  \label{eqn:PDE1}
    \mathcal{C}\ts{k}\pdv{c\ts{s,k}^0}{t} &= -\frac{1}{r\ts{k}^2}\pdv{r\ts{k}}\left(r\ts{k}^2 \pdv{c\ts{s,k}^0}{r\ts{k}}\right), \quad & \kin{n, p},\\
    \label{eqn:SPMe-electrolyte-concentration}
     \mathcal{C}\ts{e} \epsilon\ts{k}\gamma\ts{e}\pdv{c\ts{e,k}^1}{t} &= -\gamma\ts{e}\pdv{N\ts{e,k}^1}{x} + 
    \begin{cases} 
    \frac{I}{L\ts{n}}, \quad &\text{k}=\text{n}, \\ 
    0, \quad &\text{k}=\text{s}, \\ 
    -\frac{I}{L\ts{p}}, \quad &\text{k}=\text{p},
    \end{cases}
    & \kin{n, s, p}, \\ 
    \label{eqn:SPMe:electrolyte:flux}
    N\ts{e,k}^1 &= -\epsilon\ts{k}^{\text{b}} D\ts{e}(1) \pdv{c^1\ts{e,k}}{x} +     
	\begin{cases} 
	  \frac{x t^+I}{\gamma\ts{e}L\ts{n}}, \quad &\text{k}=\text{n}, \\ 
     \frac{t^+I}{\gamma\ts{e}}, \quad &\text{k}=\text{s}, \\ 
	 \frac{(1-x)t^+ I}{\gamma\ts{e} L\ts{p}}, \quad &\text{k}=\text{p},
    \end{cases} & \kin{n, s, p}. \\
      \intertext{\textbf{Boundary conditions}}  
				&\pdv{c\ts{s,k}^0}{r\ts{k}}\bigg|_{r\ts{k}=0} = 0, \quad -\frac{a\ts{k}\gamma_k}{\mathcal{C}_k} \pdv{c\ts{s,k}^0}{r\ts{k}}\bigg|_{r\ts{k}=1} = 
        \begin{cases}
		  \frac{I}{L\ts{n}} \quad &\text{k}=\text{n}, \\ 
		  -\frac{I}{L\ts{p}} \quad &\text{k}=\text{p},
        \end{cases} 
        & \kin{n, p},\\ 
    &N\ts{e,n}^1\big|_{x=0} = 0, \quad N\ts{e,p}^1\big|_{x=1}=0,  &\\ 
    &c\ts{e,n}^1|_{x=L\ts{n}}=c\ts{e,s}^1|_{x=L\ts{n}}, \quad N\ts{e,n}^1\big|_{x=L\ts{n}}=N\ts{e,s}^1\big|_{x=L\ts{n}}, & \\
    &c\ts{e,s}^1|_{x=1-L\ts{p}}=c\ts{e,p}^1|_{x=1-L\ts{p}}, \quad N\ts{e,s}^1\big|_{x=1-L\ts{p}}=N\ts{e,p}^1\big|_{x=1-L\ts{p}}. & 
      \intertext{\textbf{Initial conditions}}  
      &c\ts{s,k}^0(r\ts{k},0) = c\ts{k,$0$}, & \kin{n, p}, \\
      &c\ts{e,k}^1(x,0) = 0, & \kin{n, s, p}.
    \intertext{\textbf{Terminal voltage}}
    &V = \mybar{U}_{\text{eq}}+\mybar{\eta}_r
     +  \mybar{\eta}_c + \mybar{\Delta\Phi}_{\text{Elec}} +\mybar{\Delta\Phi}_{\text{Solid}},
     \intertext{where} 
     & \mybar{U}_{\text{eq}} = U\ts{p}(c\ts{s,p}^0)\big|_{r\ts{p}=1} - U\ts{n}(c\ts{s,n}^0)\big|_{r\ts{n}=1}, \\ 
	 &\mybar{\eta}_{r} = -2\sinh^{-1}\left(\frac{I}{\mybar{j}\ts{$0$,p} L\ts{p}}\right)
	 -2\sinh^{-1}\left(\frac{I}{\mybar{j}\ts{$0$,n} L\ts{n}}\right), \\
     \label{eqn:SPMe:concentration:overpotential}
	 &\mybar{\eta}_c =  2 \mathcal{C}\ts{e} (1-t^+)\left(\mybar{c}\ts{e,p}^1 - \mybar{c}\ts{e,n}^1\right), \\
     \label{eqn:SPMe:negative_exchange}
	 &\mybar{j}\ts{$0$,n} =  \frac{1}{L\ts{n}}\int_0^{L\ts{n}} \frac{\gamma\ts{n}}{\mathcal{C}\ts{r,n}} (c\ts{s,n}^0)^{1/2}(1-c\ts{s,n}^0)^{1/2} (1+\mathcal{C}\ts{e} c\ts{e,n}^1 )^{1/2} \, \text{d}x, \\
     \label{eqn:SPMe:positive_exchange}
	 &\mybar{j}\ts{$0$,p} =  \frac{1}{L\ts{p}}\int_{1-L\ts{p}}^{1} \frac{\gamma\ts{p}}{\mathcal{C}\ts{r,p}}  (c\ts{s,p}^0)^{1/2}(1-c\ts{s,p}^0)^{1/2} (1+\mathcal{C}\ts{e} c\ts{e,p}^1 )^{1/2} \, \text{d}x, \\
	 &\mybar{\Delta \Phi}_{\text{Elec}}= -\frac{I}{\hat{\kappa}\ts{e} \kappa\ts{e}(1)}\left(\frac{L\ts{n}}{3\epsilon\ts{n}^b} + \frac{L\ts{s}}{\epsilon\ts{s}^b} + \frac{L\ts{p}}{3\epsilon\ts{p}^b} \right),
     \label{eqn:SPMe:electrolyte_ohmic_losses} \\
	 &\mybar{\Delta \Phi}_{\text{Solid}} =  -\frac{I}{3}\left(\frac{L\ts{p}}{\sigma\ts{p}} + \frac{L\ts{n}}{\sigma\ts{n}} \right).
     \end{align} 
\end{subequations} 

Here the overbar terms are electrode-averaged quantities. The SPMe (\ref{eqn:SPMe}) consist of two independent linear PDE problems that describe the concentration of lithium in the negative and positive particles and an independent linear PDE problem that describes the concentration of lithium ions in the three regions of the electrolyte. The terminal voltage is obtained post-solution through a simple and easily interpreted algebraic expression. Since all three PDE problems are independent, the problem has a naturally parallel structure. The linearity of the PDEs is also advantageous for the application of numerical methods and the determination of simpler/analytic solutions in special cases (e.g. the SPMe(S)). 

\section{Model comparisons}\label{sec:SPM-SPMe-DFN}
\subsection{Finite-volume implementation}
In this section, we compare the DFN model (\ref{eqn:DFN}), SPM (\ref{eqn:SPM}), and the SPMe (\ref{eqn:SPMe}). We implement the DFN model by discretizing the spatial dimensions using the finite-volume method to convert the system of PDEs into a system of differential algebraic equations (DAEs) of index one. Before solving this system, a set of consistent initial conditions for the potentials are found numerically using Newton's method. The time evolution of the system is then performed using the SUNDIALs DAE solvers interfaced though PyBaMM (Python Battery Mathematical Modelling) \cite{sundials, pybamm}. PyBaMM is a battery modelling software implemented in Python which is designed to facilitate the comparison of battery models by providing a common interface to discretization methods and numerical solvers. Similarly, we use the finite-volume method to discretize the spatial dimensions of the SPM and SPMe and again use SUNDIALs for the time evolution. We use the same mesh to discretize the SPM, SPMe, and DFN model. In the $x$-direction, we use $30$ points in the negative electrode, $20$ points in the separator, and $30$ points in the positive electrode. In the $r$-direction, we use $15$ points. Numerical errors associated with the spatial discretization, are therefore of order $10^{-2}$. However, we have limited their influence upon the comparison by applying the same numerical method to each model. Since we aim to compare models and not numerical methods, we only concern ourselves with the relative reduction in computational complexity obtained by using the SPMe instead of the DFN model, whilst retaining the same numerical method. Here, we use the finite-volume method, but of course alternative methods could be applied to both models to further increase speed and/or reduce memory requirements of each. 

Computational complexity consists of space complexity (memory) and time complexity. To demonstrate the reduction in space complexity, we consider our finite-volume implementation with $30$ points in the electrodes, $20$ points in the separator, and $15$ points in the particles. In this case, the DFN model requires $2\times(30 \times 15)=900$ states for the concentration of lithium in the particles, $80$ states for the concentration of lithium ions in the electrolyte, $80$ states for the electrolyte potential, and $60$ states for the electrode potentials. This leads to a total of $1120$ internal states, which are to be stored at each time step. On the other hand, the SPMe requires only $30$ states for the concentration of lithium in the particles and $80$ states for the concentration of lithium ions in the electrolyte. Therefore, for this discretization the SPMe requires just over $10\%$ of the memory required by the DFN model. With regards to time complexity, the DFN model is limited in two respects. Firstly, it is limited by the large number of states which must be computed at every time step and secondly it is limited by the stiff nature of the system, which arises from the discretization of the mixed parabolic and elliptic PDEs resulting in a system of DAEs. As previously discussed, the SPMe addresses the first of these limitations, however, it also addresses the second limitation because the model consists of only three linear parabolic PDEs, which upon discretization lead to a well-conditioned system of ODEs amenable to taking larger time steps than the stiff DAE system of the DFN model. Furthermore, being a system of DAEs the DFN model suffers from convergence issues if inconsistent initial conditions are provided or if there is a large change in current, for example, in switching between charging and discharging. This is an inherent issue with DAE based algorithms and efforts have been made to limit the range of inconsistent initial conditions \cite{Methekar2011}. However, this robustness problem persists and convergence of the DFN model cannot always be ensured for non-constant currents. 

\subsection{Voltage comparison}
\begin{figure}[h]
	\centering
    \includegraphics{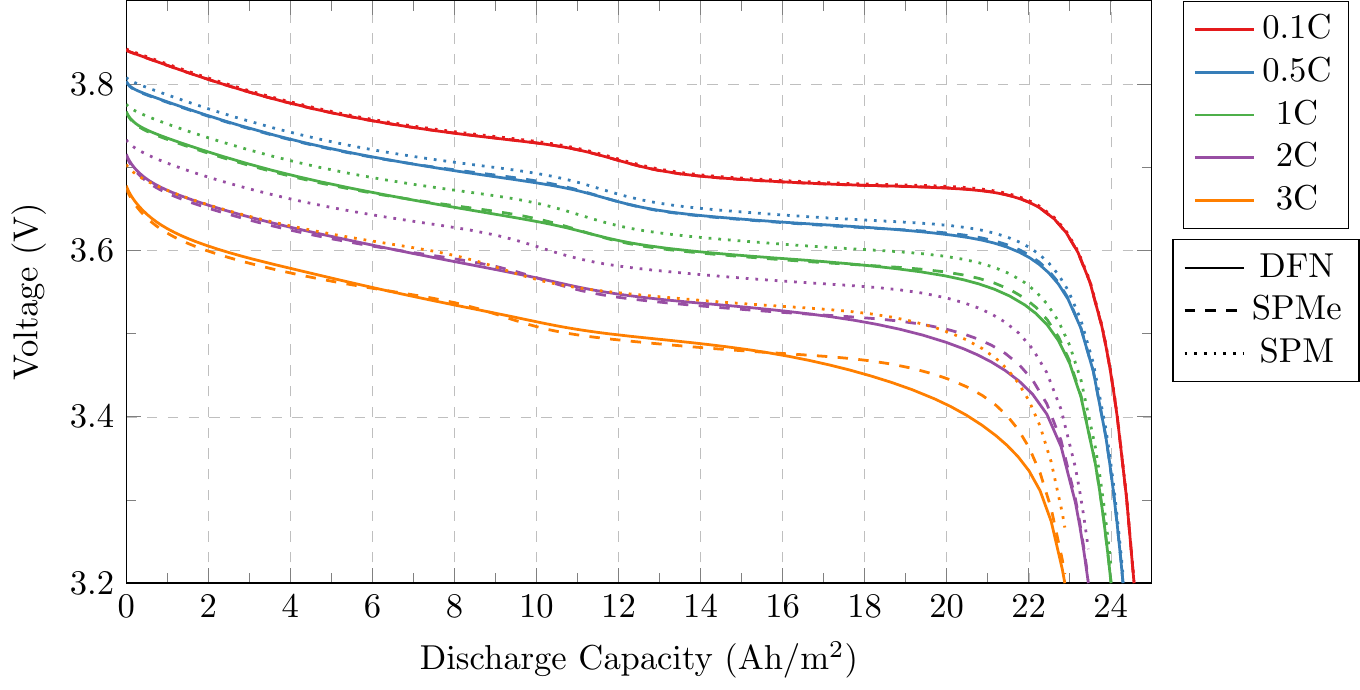}
	\caption{Constant-current discharge comparison of DFN model (\ref{eqn:DFN}), SPM (\ref{eqn:SPM}), and SPMe (\ref{eqn:SPMe}) over a range of typical C-rates. The root mean square (RMS) voltage error for the reduced models with respect to the DFN model is provided in Table~\ref{tab:RMSE-table}.} \label{fig:voltage} 
\end{figure}

We compare the SPM, SPMe, and DFN model by considering the case of a single constant-current discharge over a range of C-rates. The initial stoichiometries of the negative and positive electrodes are 0.8 and 0.6, respectively, and we terminate the discharge when the terminal voltage reaches \SI{3.2}{V}. For this cell a C-rate of \SI{1}{C} corresponds to a discharge current density of \SI{24}{A/m^2}. As provided in Table \ref{table:dimensionless_parameter_values}, we have $\mathcal{C}\ts{e}=5.1\times 10^{-3}\mathcal{C}$ where $\mathcal{C}$ is the C-rate. The predicted terminal voltage of each model is presented in Figure \ref{fig:voltage}. 

\begin{figure}[h]
    \centering
    \includegraphics{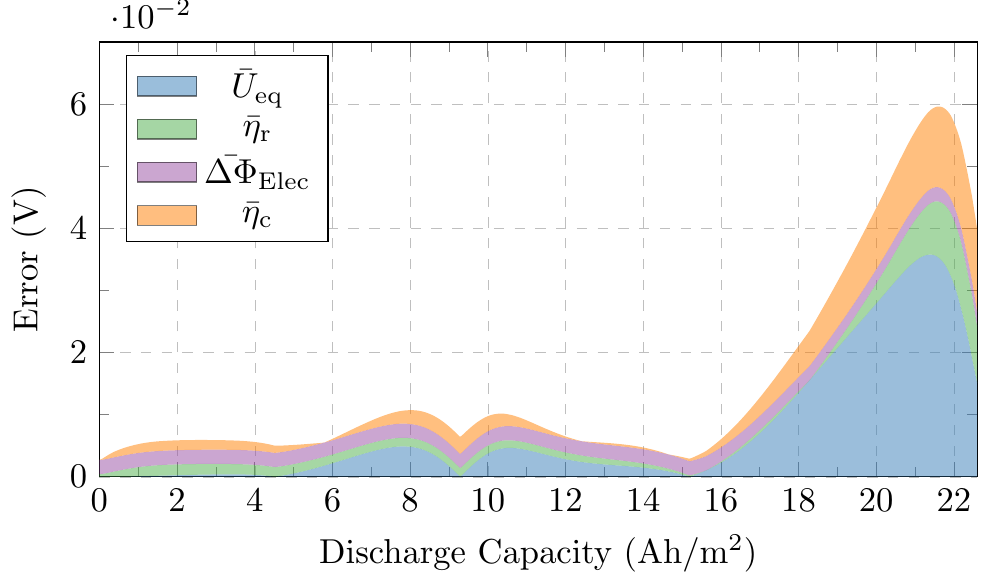}
  \caption{Overpotential errors between the SPMe and DFN.}
  \label{fig:overpotential-errors}
\end{figure}

At low C-rates, all three models match well with a RMS voltage error of just \SI{1.72}{mV} for the SPM at \SI{0.1}{C}, as expected. However, at higher C-rates we observe that the SPM prediction deviates from the DFN model solution, producing a RMS voltage error of \SI{19.86}{mV} at \SI{1}{C} and \SI{62.78}{mV} at \SI{3}{C}. The SPMe greatly improves upon this, with a RMS voltage error of just \SI{3.04}{mV} at \SI{1}{C} and \SI{13.34}{mV} at \SI{3}{C}. Unfortunately, there is a discrepancy in the voltage curves near then end of the discharge at $\SI{3}{C}$. To investigate the source of this discrepancy, we plot the error in each component of the voltage during a $\SI{3}{C}$ discharge in Figure \ref{fig:overpotential-errors}. 

We observe that for the majority of the discharge all components of the SPMe voltage agree well with the voltage predicted by the DFN model. However, near the end of the discharge, there is a large increase in the error of our solution, as observed in Figure \ref{fig:voltage}. Around 60-70\% of this error is due to a poor estimation of the electrode-averaged OCV. This error occurs when the OCV becomes highly nonlinear. If we extend our asymptotic expansion of the OCP, $U\ts{k}(c\ts{k})$ to second order, we obtain the term $\mathcal{C}\ts{e}^2 U\ts{k}''(c\ts{s,k}^0) (c\ts{s,k}^1)^2 / 2$. Hence, when the OCV is highly nonlinear, $U\ts{k}''(c\ts{s,k}^0)$ becomes large and the higher-order terms neglected by the SPMe become significant. To account for this behavior, we can consider the distinguished limit in which 
\begin{equation}\label{eqn:additional-limit}
  U\ts{k}''(c\ts{s,k}^0) \sim \mathcal{O}(\mathcal{C}\ts{e}^{-1}).
\end{equation}
In this limit, we cannot avoid solving for the concentrations in all the particles in each electrode and therefore lose a large portion of the computational simplicity of the SPM and SPMe. We have developed and implemented a numerical scheme for this limit and found that it does indeed correct the voltage discrepancy. However, because of the increased computational complexity, we do not discuss it in detail within this paper.

\subsection{Grid dependence and computation time}
\begin{figure}[h]
  \centering
  \includegraphics{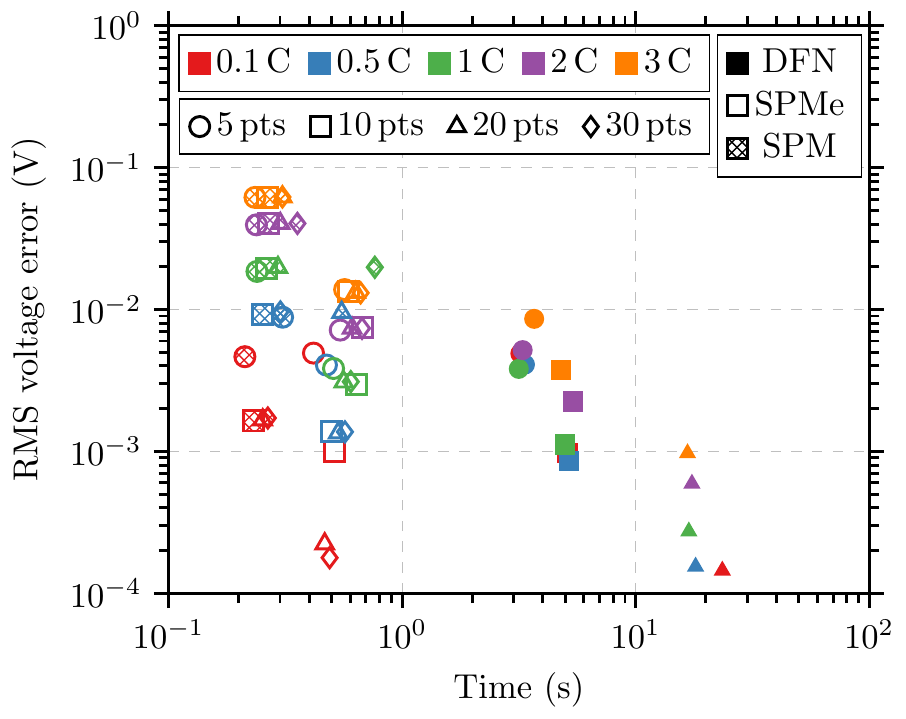}
  \caption{RMS voltage error relative to the DFN model with $30$ $x$-grid points in each of the domains: negative electrode, separator, and positive electrode; as well as $30$ $r$-grid points in the particles. This reference model with $30$ points in each domain has an average (across C-rates) run time of \SI{120}{s}. Here, `$n$~pts' refers to $n$ points in each of the domains: negative electrode, separator, and positive electrode as well as $n$ points in each of the particles.}
  \label{fig:time_v_error}
\end{figure}

In Figure~\ref{fig:time_v_error}, we compare the solutions of the SPMe and DFN model with $5$, $10$, $20$ and $30$ points in each domain (negative electrode, separator, positive electrode, negative particle, positive particle) across a range of C-rates. Here, we measure the RMS voltage error with respect to the DFN model solution with $30$ points in each domain. It is important here to only focus upon the relative timings of the models and not absolute times. Of course, a pure C++ (instead of Python) implementation of the models with a more sophisticated numerical method would increase speed of both models, but that has not been our focus. It should also be noted that this comparison is for a constant-current discharge; the non-constant case has been reported to give rise to even longer computation times for the DFN model \cite{Bizeray2015}. The key observation to make from Figure~\ref{fig:time_v_error} is that for a relatively small increase in RMS voltage error, particularly at low C-rates, an order of magnitude decrease in computation time is achieved by using the SPMe instead of the DFN model. Achieving such large decreases in computation time is consistent across all C-rates and numbers of grid points used. Further, we can observe that the SPMe generally increases the accuracy of the SPM by an order of magnitude for a particular C-rate whilst maintaining a similar computation time. When a small number of grid points (e.g. 10 points) are used, and a current with a C-rate above \SI{0.5}{C} is applied, the SPMe produces a solution that is not limited by the discretization error; we discuss the limiting asymptotic errors in Section~\ref{sec:dimensional-SPMe}. Therefore, to achieve the asymptotic accuracy of the SPMe, a coarser spatial discretization is often sufficient, which could also allow for further increases in speed and savings in memory.

\subsection{Comparison of internal states}
\begin{figure}[h]
  \centering
  \includegraphics{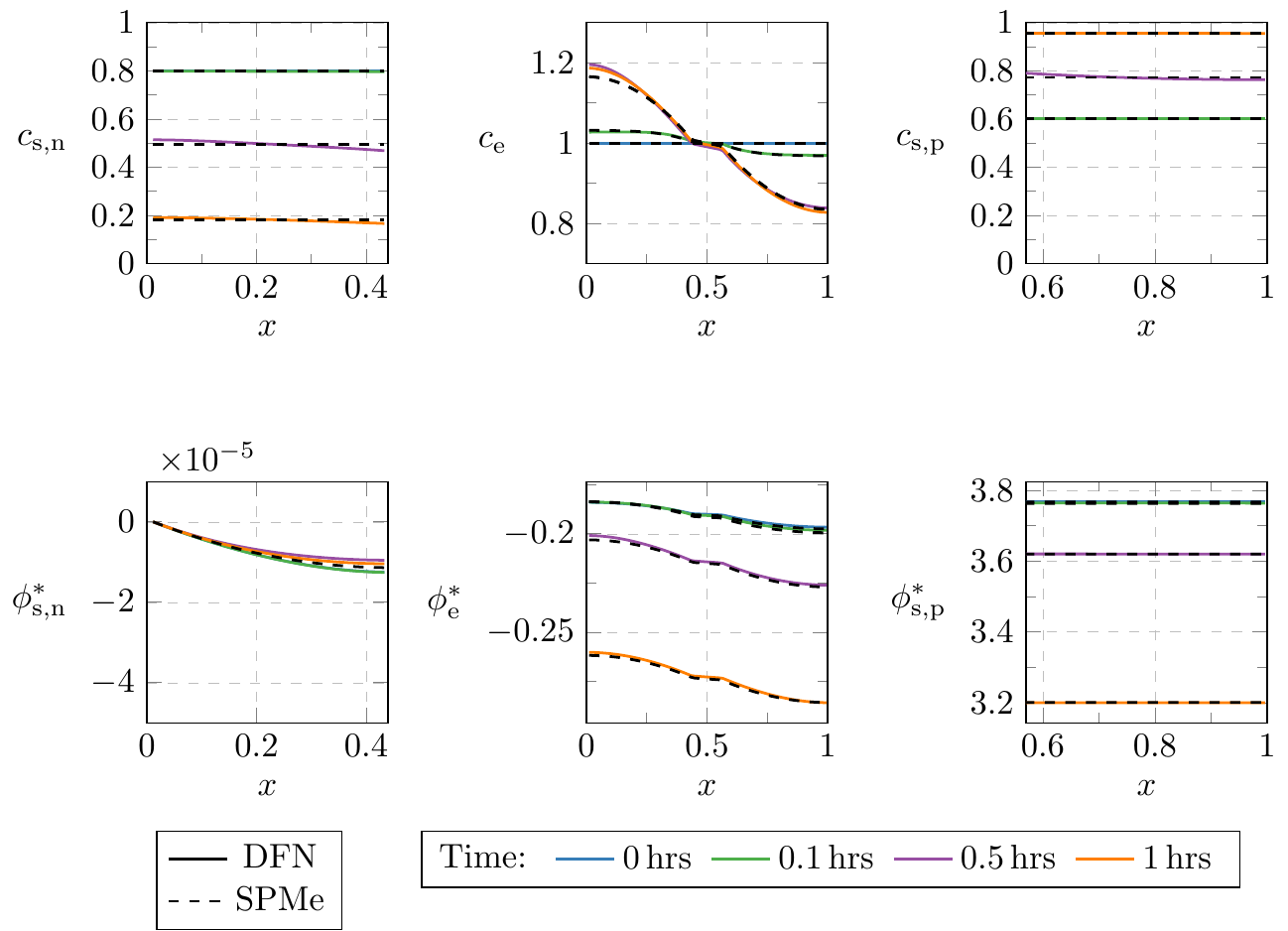}
  \caption{Comparison of DFN model and SPMe internal states during a $\SI{1}{C}$ constant current discharge. The DFN model solution is given by the solid lines and the SPMe solution by the closest black dashed line. Note that some of the black dashed lines lie upon others so it appears that there are fewer black dashed lines than solid lines.}
  \label{fig:internal-states}
\end{figure}

To further confirm the accuracy of the SPMe, we compare the internal states predicted by the SPMe and DFN model. These are presented for a $\SI{1}{C}$ constant current discharge in Figure \ref{fig:internal-states}. We observe good agreement between the two sets of model predictions. However, two key discrepancies can be observed: the first in the negative electrode stoichiometry and the second in the electrolyte concentration at late times. We note that the apparent discrepancy in the negative electrode potential is only due to the scale employed in Figure \ref{fig:internal-states}. This discrepancy is in fact $\mathcal{O}(\mathcal{C}\ts{e}^2)$. 

The discrepancy in the negative electrode stoichiometry is a result of the SPMe using only the leading-order equations in $\mathcal{C}\ts{e}$ within the electrode particles; this is equivalent to considering only the electrode-averaged concentration in the particles. This is the same approximation as employed by the SPM and the SPMe is therefore only accurate to $\mathcal{O}(\mathcal{C}\ts{e})$ for the concentration in the particles. It is possible to additionally solve for the first-order correction in the particles, but this requires solving a diffusion equation in each particle instead of a single diffusion equation in the electrode-averaged particle. This increases the computational complexity of the model and has therefore been omitted from this paper.

The discrepancy in the electrolyte concentration at late times is fundamentally connected to the discrepancy in the voltage curves observed in Figure \ref{fig:voltage}. As already discussed, this is due to nonlinear nature of the OCV and the term $U\ts{k}''(c\ts{s,k}^0)$ becoming large. We must then consider the distinguished limit in which (\ref{eqn:additional-limit}) holds. In this limit, we consider a heterogeneous interfacial current density, and so the electrode-averaged current density source/sink terms in (\ref{eqn:SPMe-electrolyte-concentration}) and (\ref{eqn:SPMe:electrolyte:flux}) are replaced by heterogeneous versions. We have implemented a numerical scheme for this limit and can confirm that this discrepancy is accounted for in this way. However, this limit requires one to solve for the concentration in each of the particles and is therefore much more computationally expensive than the models of concern in this paper.

\begin{figure}[h]
  \centering
  \includegraphics{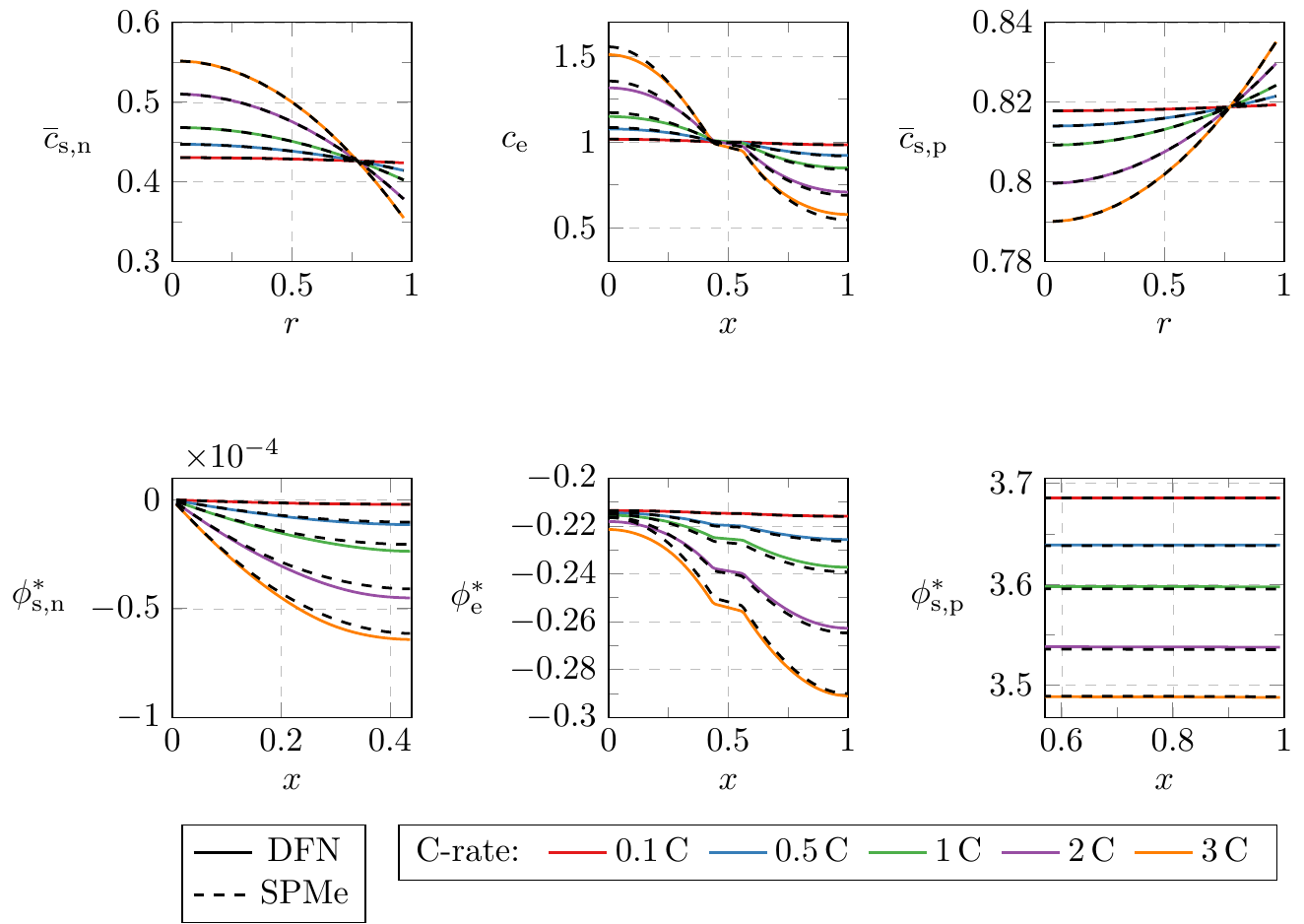}
  \caption{Comparison of internal state across multiple C-rates at a discharge capacity of \SI{15}{Ah/m^2}. The DFN model solution is given by the solid lines and the SPMe solution by the closest black dashed line. Note that we plot the $x$-averaged particle concentration here instead of the surface concentration plotted in Figure~\ref{fig:internal-states}.}
  \label{fig:c-rate-internal}
\end{figure}

In Figure~\ref{fig:c-rate-internal}, we compare the internal states of the SPMe and DFN model across a range of C-rates. Here, we display the lithium concentration in the x-averaged particle to demonstrate the ability of our model to capture the concentration profile inside particles as well as the surface concentration. As before, we observe good agreement of the internal states, with the exception of the electrolyte concentrations and the electrolyte potentials at large C-rates. Both of these discrepancies are a result of the the nonlinearity of the OCV, as discussed previously.

\subsection{Alternative numerical methods}
In the previous section, we only discussed the benefits of the SPMe compared to the DFN model when both models were implemented using the finite-volume method. However, there are other more sophisticated methods that have been applied to the DFN model, which can also be applied to the SPMe (e.g. \cite{Northrop2011a}). The approach implemented in \cite{Northrop2011a} involves volume averaging the PDEs in the particles and employing a polynomial approximation to obtain the surface concentration. As a result, in their formulation of the DFN model (Table IV of \cite{Northrop2011a}), there are 4 PDEs, and 1 algebraic equation in each electrode along with 2 PDEs in the separator to be solved. Upon applying their orthogonal collocation discretization procedure, $5(N\ts{n} + 1) + 2(N\ts{s} + 1) + 5(N\ts{p} + 1)$ DAEs are obtained (see just after equation (26) of \cite{Northrop2011a} for more details), where $N\ts{k}$ is the number of cosine basis functions used in each domain. If we were to reformulate the governing equations in the particles in the SPMe using their polynomial approximation approach, we would have 1 PDE for the electrolyte concentration and 1 ODE for the $r$-averaged lithium concentration in the average particle in each electrode (note that the algebraic constraint for the surface concentration can now be solved analytically) and 1 PDE for the electrolyte concentration in the separator. Therefore, upon application of their discretization, we obtain $((N\ts{n} + 1) + 1) + (N\ts{s} + 1) + ((N\ts{p} + 1) + 1)$ ODEs. Using a single cosine in each domain results in $8$ ODEs for the SPMe compared to $24$ DAEs for the DFN model. Alternatively, using $N\ts{n}=7$, $N\ts{s}=3$, and $N\ts{p}=7$ produces a system of 88 DAEs for the DFN model but only $22$ ODEs for the SPMe. Therefore, applying this discretization to the SPMe instead of the DFN model can result in the memory requirements being reduced to below a third. This is a step towards employing a practical implementation of a physics-based model in a BMS where RAM is limited. Alternatively, a system of battery packs consisting of more than three times as many cells can be solved with similar RAM requirements using the SPMe instead of the DFN model. In addition, it is also typically faster to solve fewer ODEs (two of which are independent of the others) than a greater number of highly-coupled DAEs. Furthermore, the convergence issues related to DAE systems are not an issue for the ODE system produced by discretizing the SPMe. These types of benefits will be consistent across different discretization approaches.

To provide an example of the benefits of using the SPMe with the discretization in \cite{Northrop2011a} instead of the DFN model, we consider the case where there is a memory limit such that only $24$ states can be stored at each time step. In this case, the DFN model is restricted to employing just one cosine basis function in each domain, whereas the SPMe can make use of $N\ts{n}=7$, $N\ts{s}=3$, $N\ts{p}=7$ basis functions. The RMS voltage error for a \SI{1}{C} discharge for the DFN model using a single cosine function in each domain relative to a full finite difference implementation (with 50 points in each electrode and 35 points in the separator) is \SI{17.84}{mV}, as reported in \cite{Northrop2011a}. In \cite{Northrop2011a}, using $N\ts{n}=7$, $N\ts{s}=3$, $N\ts{p}=7$ basis functions, the RMS voltage error for the DFN model is just \SI{0.91}{mV}. If we assume that this numerical error carries across to an SPMe implementation (this appears to be the case in Figure \ref{fig:time_v_error}, for example for \SI{0.1}{C} with \SI{5}{pts}), then the main source of error will be modelling error. For a \SI{1}{C} discharge, the SPMe produces an error of just \SI{3.04}{mV} in the finite-volume implementation, see Table~\ref{tab:RMSE-table} (this RMS voltage error also holds for 50 points in the electrodes and 35 points in the separator). Assuming that this error is entirely modelling error, then we expect the RMS voltage error of a $N\ts{n}=7$, $N\ts{s}=3$, $N\ts{p}=7$ discretization of the SPMe to be approximately \SI{3.04}{mV}. Therefore, with the same memory budget employing the SPMe instead of the DFN model can improve the accuracy of the voltage prediction by an order of magnitude.

\section{Critical assessment of variations on the SPMe in the literature} 
\subsection{Overview of models}
There are a number of alternative models in the literature that extend the SPM in an ad-hoc manner to account for electrolyte effects \cite{kemper2013extended, Perez2016, prada2012simplified, han2015simplification, rahimian2013extension, tanim2015temperature}. In this section, we highlight the key differences between these models and the canonical SPMe (\ref{eqn:SPMe}) presented here. We have chosen to compare a subset of the models, which cover the variety of ad-hoc models available. To do this, we have converted the models to dimensionless form using the scalings in (\ref{eqn:scalings}). In some papers the model is discretized during development. We view the choice of discretization to be a numerical method instead of a feature of the model itself. Therefore, we have converted each model into continuum form so as to only highlight the differences in the underlying models. We do not aim to study the benefits and drawbacks of different numerical methods.

A common theme in the models from the literature is to replace the electrode-averaged concentration overpotential and electrode-averaged electrolyte Ohmic losses with pointwise versions. It is also common to neglect the solid-phase Ohmic losses (this is a reasonable assumption since we have already observed these to be small). The general form of the terminal voltage expression used in the literature is then: 
\begin{equation} \label{eqn:literature-pointwise-voltage}
  V = \mybar{U}\ts{eq} + \mybar{\eta}\ts{r} + \eta\ts{c}\big|_{x\ts{n}=0,x\ts{p}=1} + \Delta\Phi\ts{Elec}\big|_{x\ts{n}=0, x\ts{p}=1}.
\end{equation}
This expression consists of a combination of both electrode-averaged and pointwise terms, and therefore an accuracy of $\mathcal{O}(\mathcal{C}\ts{e}^2)$ cannot be ensured.

We begin by considering the model proposed by Perez et.~al. in \cite{Perez2016}. Firstly, the electrolyte flux (\ref{eqn:SPMe:electrolyte:flux}) is replaced by the expression (after converting to our notation)
\begin{equation}\label{eqn:alternative_flux}
  N\ts{e,k}^1 = -\epsilon\ts{k}^bD\ts{e}(1+\mathcal{C}\ts{e} c\ts{e,k}^1)\pdv{c\ts{e,k}^1}{x} + 
	\begin{cases} 
	  \frac{x t^+I}{\gamma\ts{e}L\ts{n}}, \quad &\text{k}=\text{n}, \\ 
     \frac{t^+I}{\gamma\ts{e}}, \quad &\text{k}=\text{s}, \\ 
	 \frac{(1-x)t^+ I}{\gamma\ts{e} L\ts{p}}, \quad &\text{k}=\text{p},
    \end{cases}
	\quad \text{for } \kin{n, s, p}.
\end{equation}
Due to the the presence of the nonlinear diffusion coefficient, the electrolyte problem is nonlinear, while that for the SPMe (\ref{eqn:SPMe}) is linear. Since the other terms are unchanged, this nonlinear form is also accurate up to $\mathcal{O}(\mathcal{C}\ts{e}^2)$. Equation (\ref{eqn:alternative_flux}) includes some (but not all) higher-order terms, which in practice may increase accuracy but this cannot be ensured (there is also a chance it could reduce accuracy). Secondly, the electrode-averaged concentration overpotential (\ref{eqn:SPMe:concentration:overpotential}) and the electrode-averaged electrolyte Ohmic losses (\ref{eqn:SPMe:electrolyte_ohmic_losses}) are replaced by their pointwise versions:
\begin{align}
    \label{eqn:ScottMoura:concentrationOverpotential}
	&\eta\ts{c}\big|_{x\ts{n}=0,x\ts{p}=1} = 2 (1-t^+)\log\left(\frac{1+\mathcal{C}\ts{e} c\ts{e,p}^1\big|_{x=1}}{1+\mathcal{C}\ts{e} c\ts{e,p}^1\big|_{x=0}} \right), \\
	&\Delta \Phi_{\text{Elec}}\big|_{x\ts{n}=0,x\ts{p}=1}  = -\frac{I}{2\hat{\kappa}\ts{e} \mybar{\kappa}\ts{e}^{\text{eff}}}\left(L\ts{n} + 2L\ts{s} + L\ts{p} \right), \label{eqn:ScottMoura:ohmicLosses}
\end{align}
respectively (note that to get (\ref{eqn:ScottMoura:ohmicLosses}) we have corrected the sign of the expression in \cite{Perez2016}). Here, $\mybar{\kappa}\ts{e}^{\text{eff}}$ is the dimensionless effective conductivity averaged across the entire cell, with the effective conductivity defined by $\kappa\ts{e}^{\text{eff}}(c\ts{e,k})=\epsilon\ts{k}^b\kappa\ts{e}(c\ts{e,k})$. The terminal voltage is then given by (\ref{eqn:literature-pointwise-voltage}). Additionally, (\ref{eqn:ScottMoura:ohmicLosses}) requires that $\kappa_e^{\text{eff}}(c\ts{e,k})\approx\mybar{\kappa}_e^{\text{eff}}$ throughout the cell. With this assumption, $\mathcal{O}(\mathcal{C}\ts{e}^2)$ accuracy cannot be ensured for all values of $\epsilon\ts{n}$, $\epsilon\ts{s}$, and $\epsilon\ts{p}$. Finally, solid-phase Ohmic losses are neglected and Ohmic losses due to the presence of SEI are included; we shall neglect the SEI terms in our comparisons, noting that (\ref{eqn:SPMe}) can be easily extended to include them.

The model presented by Prada et.~al. \cite{prada2012simplified} also 
employs (\ref{eqn:ScottMoura:concentrationOverpotential}) for the concentration
overpotential. However, the electrolyte Ohmic losses are taken to be 
\begin{equation}\label{eqn:Prada}
  \Delta \Phi_{\text{Elec}}\big|_{x\ts{n}=0,x\ts{p}=1} = -\frac{I}{2\hat{\kappa}\ts{e}} \left(\frac{L\ts{n}}{\epsilon\ts{n}^b\mybar{\kappa}\ts{e,n}} + 2\frac{L_s}{\epsilon\ts{s}^b \mybar{\kappa}\ts{e,s}} + \frac{L\ts{p}}{\epsilon\ts{p}^b\mybar{\kappa}\ts{e,p}} \right), 
\end{equation}
where
\begin{equation}\label{eqn:sigma_bar_notation}
  \begin{aligned}
    &\mybar{\kappa}\ts{e,n} = \int_{0}^{L\ts{n}} \kappa\ts{e}\left(1+\mathcal{C}\ts{e} c\ts{e,n}^1\right)\,\text{d}x, \\
    &\mybar{\kappa}\ts{e,s} = \int_{L\ts{n}}^{1-L\ts{p}} \kappa\ts{e}\left(1+\mathcal{C}\ts{e} c\ts{e,s}^1\right)\,\text{d}x, \\
    &\mybar{\kappa}_{e,p} = \int_{1-L\ts{p}}^1 \kappa\ts{e}\left(1+\mathcal{C}\ts{e} c\ts{e,p}^1\right)\,\text{d}x.
  \end{aligned}
\end{equation}
Whilst (\ref{eqn:Prada}) does not rely upon the assumption that 
$\kappa\ts{e}^{\text{eff}}(c\ts{e,k})\approx\mybar{\kappa}\ts{e}^{\text{eff}}$, 
its form is still a result of considering the pointwise electrolyte potential difference instead of the 
electrode-averaged difference. In addition to these differences, Prada et.~al. \cite{prada2012simplified} take the exchange-current densities $\mybar{j}\ts{0,n}$ and $\mybar{j}\ts{0,p}$ to be constant. In terms of accurately reproducing the results of the DFN model, this simplification has a clear disadvantage as the reaction overpotentials are strong functions of the lithium and lithium-ion concentrations. 

The model developed by Han et.~al \cite{han2015simplification} is the same as that presented by Prada et.~al. \cite{prada2012simplified} without the additional assumption of constant exchange-current densities. That is, Han et.~al. \cite{han2015simplification} employ (\ref{eqn:ScottMoura:concentrationOverpotential}) and (\ref{eqn:Prada}), which are the pointwise concentration overpotential, and electrolyte Ohmic losses, respectively. Han et.~al. \cite{han2015simplification} note the tendency for their model to over-correct the voltage when accounting for electrolyte effects. We suspect the use of pointwise terms is the cause.

The model presented by Kemper et.~al. \cite{kemper2013extended} is somewhat different to the others we have discussed. Firstly, the model is presented as a set of ODEs instead of PDEs. These ODEs are derived by spatially discretizing the underlying PDEs. Whilst this particular discretization may be useful, we consider this to be a numerical method and not a feature of the model itself. Since we aim to compare the underlying simplified models directly, we have converted these ODEs back into PDEs. The resulting PDEs that describe the concentrations in the electrode particles and the electrolyte, are equivalent to those used in our model. However, the expression for the terminal voltage is very different and it is not easy to prescribe meaning to each of the individual components; we have however attempted this. We have converted the voltage from \cite{kemper2013extended} into dimensionless form and provided details in Appendix \ref{app:dimVoltage} to be clear about the exact model we are comparing. It was unclear if the components of the voltage correspond to electrode-averaged or pointwise quantities so we have left this unstated. 

\subsection{Model comparison}
\begin{figure}[h]
  \centering
  \includegraphics{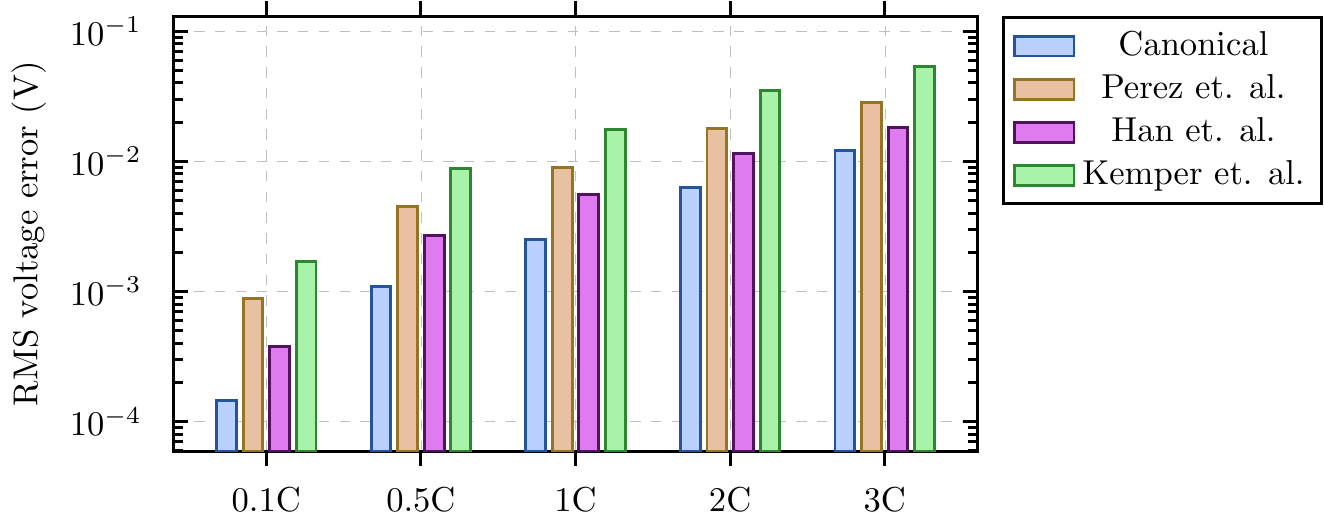}
  \caption{Comparison of versions of the SPMe: Canonical (\ref{eqn:SPMe}), Perez et. al. \cite{Perez2016}, Han et. al. \cite{han2015simplification}, and Kemper et. al. \cite{kemper2013extended}. The models are compared by considering the RMS voltage error of the simplified model voltage prediction vs the DFN model (\ref{eqn:DFN}).}
  \label{fig:model-variations}
\end{figure}

We compare the variations of the SPMe in the literature and our canonical SPMe. For this section, we use a finite-volume implementation of each of the models implemented in MATLAB with ODE15s being used for the time integration. We consider a range of constant-current discharge rates and then consider the RMS voltage error of each model relative to the DFN model. For each model, we discretize using 30 points in each electrode, 20 in the separator, and 15 in each particle. Our results are presented in Figure \ref{fig:model-variations}, where we compare the models in \cite{Perez2016, han2015simplification, kemper2013extended}. Each version of the SPMe consists of three parabolic PDEs, one in the negative particle, one in the positive particle, and one in the electrolyte alongside an algebraic expression for the voltage. As a result, each SPMe must store $2\times15 + (30 + 20 + 30)=120$ states at each time step. Therefore, the memory requirements of each model are similar. Upon evaluation, each version of the model takes on average \SI{0.07}{s} (at the time of writing the ODE15s implementation is faster the Python--SUNDIALs implementation). We observe that across all discharge rates, our canonical SPMe outperforms the other models from the literature. In particular, our canonical SPMe is consistently an order of magnitude more accurate than the models in Perez et.~al. \cite{Perez2016} and Kemper et.~al \cite{kemper2013extended}. Furthermore, at higher C-rates, the RMS voltage errors in the models from the literature approach being of the order of $\SI{0.1}{V}$ whereas the RMS voltage errors of our SPMe only reach the order of $\SI{0.01}{V}$. Additionally, as we would expect, our model converges to the DFN model solution at a faster rate than the other models. We attribute the main gains of our model to the consistent electrode-averaged OCPs, overpotentials, and Ohmic losses in our terminal voltage expression.

\section{Dimensional model summary and conditions for application}\label{sec:dimensional-SPMe}
We now present a summary of our dimensional SPMe alongside the conditions that should be met for the model to be valid. The purpose of this section is to serve as a reference, which is easily accessible to those not interested in the full details of the model derivation. For the purpose of this section, we will drop the superscript notation, which was used to indicated the asymptotic order of a variable. Also, to be consistent with the rest of the paper, we employ a superscript $*$ to denote dimensional quantities. By reapplying our scalings in (\ref{eqn:scalings}) to (\ref{eqn:SPMe}) and combining the leading- and first-order equations for the electrolyte concentration, we obtain:
\begin{subequations}\label{eqn:SPMe-dimensional}
    \begin{align}
    \intertext{\textbf{Governing Equations}}
    \pdv{c\ts{s,k}^*}{t^*} &= \frac{1}{(r^*)^2}\pdv{r^*}\left((r^*)^2 D\ts{s,k}^*(c\ts{s,k}^*)\pdv{c\ts{s,k}^*}{r^*}\right),  \quad  \kin{n, p},\\
    \label{eqn:SPMe-dimensional-electrolyte}
     \epsilon\ts{k}\pdv{c\ts{e,k}^*}{t^*} &= -\pdv{N\ts{e,k}^*}{x^*} + 
    \begin{cases} 
    \frac{I^*}{F^*L\ts{n}^*}, \quad &\text{k}=\text{n}, \\ 
    0, \quad &\text{k}=\text{s}, \\ 
    -\frac{I^*}{F^*L\ts{p}^*}, \quad &\text{k}=\text{p},
    \end{cases}
    \quad \kin{n, s, p}, \\ 
    N\ts{e,k}^* &= -\epsilon\ts{k}^{\text{b}} D\ts{e}^*(c\ts{e,typ}^*) \pdv{c^*\ts{e,k}}{x} +     
	\begin{cases} 
	  \frac{x^* t^+I^*}{F^*L\ts{n}^*}, \quad &\text{k}=\text{n}, \\ 
     \frac{t^+I^*}{F^*}, \quad &\text{k}=\text{s}, \\ 
	 \frac{(L^*-x^*)t^+ I^*}{F^* L\ts{p}}, \quad &\text{k}=\text{p},
    \end{cases} \quad \kin{n, s, p}.
      \intertext{\textbf{Boundary Conditions}}  
				&\pdv{c\ts{s,k}^*}{r^*}\bigg|_{r^*=0} = 0, \quad -D\ts{s,k}^*(c\ts{s,k}^*)\pdv{c\ts{s,k}^*}{r^*}\bigg|_{r^*=R\ts{k}^*} = 
        \begin{cases}
		  \frac{I^*}{F^* a\ts{n}^* L\ts{n}^*}, \quad &\text{k}=\text{n}, \\ 
		  -\frac{I^*}{F^* a\ts{p}^* L\ts{p}^*}, \quad &\text{k}=\text{p},
        \end{cases} 
        \quad \kin{n, p},\\ 
    &N\ts{e,n}^*\big|_{x^*=0} = 0, \quad N\ts{e,p}^*\big|_{x^*=L^*}=0,  \\ 
    &c\ts{e,n}^*|_{x^*=L\ts{n}^*}=c\ts{e,s}^*|_{x^*=L\ts{n}^*}, \quad N\ts{e,n}^*\big|_{x^*=L\ts{n}^*}=N\ts{e,s}^*\big|_{x^*=L\ts{n}^*}, \\
    &c\ts{e,s}^*|_{x^*=L^*-L\ts{p}^*}=c\ts{e,p}^*|_{x^*=L^*-L\ts{p}^*}, \quad N\ts{e,s}^*\big|_{x^*=L^*-L\ts{p}^*}=N\ts{e,p}^*\big|_{x^*=L^*-L\ts{p}^*}. 
      \intertext{\textbf{Initial Conditions}}  
      &c\ts{s,k}^*(r^*,0) = c\ts{k,$0$}^*, \quad \kin{n, p}, \\
      &c\ts{e,k}^*(x^*,0) = c\ts{e,typ}^*, \quad \kin{n, s, p}.
    \intertext{\textbf{Terminal Voltage}}
    &V^* = \mybar{U}_{\text{eq}}^*+\mybar{\eta}_r^*
     +  \mybar{\eta}_c^* + \mybar{\Delta\Phi}_{\text{Elec}}^* +\mybar{\Delta\Phi}_{\text{Solid}}^*,
     \intertext{where} 
      & \mybar{U}_{\text{eq}}^* = U\ts{p}^*\left(c\ts{s,p}^*\big|_{r^*=R\ts{p}^*}\right) - U\ts{n}^*\left(c\ts{s,n}^*\big|_{r^*=R\ts{n}^*}\right), \\ 
    &\mybar{\eta}_{r}^* = -\frac{2R^*T^*}{F^*}\sinh^{-1}\left(\frac{I^*}{a\ts{p}^*\mybar{j}\ts{$0$,p}^* L\ts{p}^*}\right)
	 -\frac{2R^*T^*}{F^*}\sinh^{-1}\left(\frac{I^*}{a\ts{n}^*\mybar{j}\ts{$0$,n}^* L\ts{n}^*}\right), \\
	 &\mybar{\eta}_c^* =  \frac{2R^*T^*}{F^* c\ts{e,typ}^*} (1-t^+)\left(\mybar{c}\ts{e,p}^* - \mybar{c}\ts{e,n}^*\right), \\
	 &\mybar{j}\ts{$0$,n}^* =  \frac{1}{L\ts{n}^*}\int_0^{L\ts{n}^*} m\ts{n}^* (c\ts{s,n}^*)^{1/2}(c\ts{s,n,max}^*-c\ts{s,n}^*)^{1/2} (c\ts{e,n}^*)^{1/2} \, \text{d}x^*, \\
	 &\mybar{j}\ts{$0$,p}^* =  \frac{1}{L\ts{p}^*}\int_{L^*-L\ts{p}^*}^{L^*} m\ts{p}^*  (c\ts{s,p}^*)^{1/2}(c\ts{s,p,max}^*-c\ts{s,p}^*)^{1/2} (c\ts{e,p}^*)^{1/2} \, \text{d}x^*, \\
   &\mybar{\Delta \Phi}_{\text{Elec}}^*= -\frac{I^*}{\kappa\ts{e}^*(c\ts{e,typ}^*)}\left(\frac{L\ts{n}^*}{3\epsilon\ts{n}^b} + \frac{L\ts{s}^*}{\epsilon\ts{s}^b} + \frac{L\ts{p}^*}{3\epsilon\ts{p}^b} \right),\\
	 &\mybar{\Delta \Phi}_{\text{Solid}}^* =  -\frac{I^*}{3}\left(\frac{L\ts{p}^*}{\sigma\ts{p}^*} + \frac{L\ts{n}^*}{\sigma\ts{n}^*} \right).
     \end{align} 
\end{subequations} 
We have provided the conditions that ensure the validity of (\ref{eqn:SPMe-dimensional}) in Table~\ref{tab:conditions}. 

If the conditions in Table~\ref{tab:conditions} are met then the model error at a particular time is of size
\begin{gather}
  \text{max}\left(
      \left(\frac{I\ts{typ}^* L^*}{D\ts{e,typ}^*F^*c\ts{n,max}^*}\right)^2, \left(\frac{I\ts{typ}^* L^*}{D\ts{e,typ}^*}\right)^2 \left(\frac{1}{F^*R^*T^*} \right)\big|(U\ts{k}^*)''\big| \right)
\end{gather}
where the term $\big|(U\ts{k}^*)''\big|$ is the absolute value of the second derivative of the OCP in electrode $\text{k}$. This term is responsible for the decrease in accuracy when the OCP is significantly nonlinear, as discussed in Section~\ref{sec:SPM-SPMe-DFN}. In situations where transient effects are not important (for example when the current varies on timescales longer than the electrolyte diffusion timescale), the same degree of accuracy can be achieved by employing the dimensional equivalent to the SPMe(S); this is achieved by neglecting the time derivative term in the electrolyte equation (\ref{eqn:SPMe-dimensional-electrolyte}).

\section{Summary and further work} 
We have systematically derived simplified mathematical models from the standard DFN model through the use of asymptotic methods. The leading-order model, the SPM (\ref{eqn:SPM}) is commonly used in the control community. By considering higher-order effects, we have extended this model to develop a canonical SPMe that is more accurate and applicable over a larger range of operating conditions. The canonical SPMe has been shown to give good agreement with the DFN model whilst providing dramatic decreases in computational complexity in the form of both reduced memory requirements and reduced computation time, both of which are highly desirable features for BMS, parameter estimation, and optimization. We have also shown our version of the SPMe be more accurate than the other reduced models of similar computational complexity available in the literature. A key result of this paper is to identify the requirement of writing the output voltage expression in terms of the electrode-averaged OCV, overpotentials, and Ohmic losses. This step has been overlooked in previous literature. Finally, our systematic approach has allowed us to identify the reasons for discrepancies in the predictions of the SPMe and identify the minimal extensions required for them to be corrected.  

There are a number of possible additional physical mechanisms that are of interest to incorporate into a battery model. These include mechanical effects, thermal effects, and degradation mechanisms. One approach would be to simply introduce these effects in an ad-hoc manner to existing simple models without consideration of their interactions within the context of a more complicated model such as the DFN model. However, as demonstrated here, it is important to derive reduced-order models in a systematic fashion; not doing so can lead to inconsistent terms and unnecessary loss of accuracy. The approach in this paper can be applied to models that include additional physical effects, ensuring that the resulting reduced-order models retain the most important terms to faithfully represent the underlying physics. Further, by conducting asymptotic analysis, the error introduced in developing simplified models can be properly quantified in terms of parameter groupings, allowing for the applicability of a model to be determined a-priori.

\appendix

\section{Electrolyte constants of integration}\label{app:electrolyte-constants}

\begin{equation}
  \phi\ts{e}^0 = - U\ts{n}(c\ts{s,n}^0\big|_{r\ts{n}=1}) - \eta\ts{n}^0
\end{equation}

\begin{equation}
  \tilde{\phi}\ts{e} = - 2(1 + t^+) \mybar{c}\ts{e,n}^1 + \frac{I L\ts{n}}{\gamma\ts{e} \kappa\ts{e}(1)} \left(\frac{1}{3\epsilon\ts{n}^b} - \frac{1}{\epsilon\ts{s}^b}\right)
  + \mybar{\phi}\ts{s,n}^1 - \mybar{\eta}\ts{n}^1
\end{equation}

\section{Electrode-averaged quantities}\label{app:electrode_averaged} 
\begin{align} 	
    \label{eqn:av-c_neg}
	&\mybar{c}\ts{e,n}^1 = \frac{(1-t^+) I}{ 6 \gamma\ts{e} D\ts{e}(1)}\left( 2\left(\frac{L\ts{p}^2}{\epsilon\ts{p}^b} - \frac{L\ts{n}^2}{\epsilon\ts{n}^b} \right) + \frac{2 L\ts{n}}{\epsilon\ts{n}^b} + \frac{3L\ts{s}}{\epsilon\ts{s}^{b}}(L\ts{p}-L\ts{n}+1)\right) \\ 
	\label{eqn:av-c_pos}
    &\mybar{c}\ts{e,p}^1 = \frac{(1-t^+) I}{ 6 \gamma\ts{e} D\ts{e}(1)}\left( 2\left(\frac{L\ts{p}^2}{\epsilon\ts{p}^b} - \frac{L\ts{n}^2}{\epsilon\ts{n}^b} \right) - \frac{2 L\ts{p}}{\epsilon\ts{p}^b} + \frac{3L\ts{s}}{\epsilon\ts{s}^{b}}(L\ts{p}-L\ts{n}-1)\right) \\ 
    &\mybar{\phi}\ts{e,n}^1 = \tilde{\phi}\ts{e} +2(1-t^+) \mybar{c}\ts{e,n}^1 + \frac{ IL\ts{n}}{\hat{\kappa}\ts{e}'\kappa\ts{e}(1)}\left(\frac{1}{3 \epsilon\ts{n}^b} - \frac{1}{\epsilon\ts{s}^b}\right) \\ 
    &\mybar{\phi}\ts{e,p}^1 = \tilde{\phi}\ts{e} +2(1-t^+) \mybar{c}\ts{e,p}^1 + \frac{ IL\ts{p}}{\hat{\kappa}\ts{e}'\kappa\ts{e}(1)}\left(\frac{1}{ \epsilon\ts{s}^b} - \frac{1}{3\epsilon\ts{p}^b}\right) - \frac{I}{\hat{\kappa}\ts{e}'\kappa\ts{e}(1)\epsilon\ts{s}^b} \\
    & \mybar{\phi}\ts{s,n}^1 = -\frac{IL\ts{n}}{3\sigma\ts{n}'} \\
    &\mybar{\varphi}\ts{s,p}^1 = \frac{IL\ts{p}}{3\sigma\ts{p}'} \\
    &\mybar{\phi}\ts{s,p}^1 = \mybar{\varphi}\ts{s,p} + V^1
\end{align}

\section{Dimensionless voltage from \cite{kemper2013extended}}\label{app:dimVoltage}
Although the form of the voltage in \cite{kemper2013extended} is not given explicitly, we assume that it is given by $\phi\ts{s}(0^+,t)-\phi\ts{s}(0^-,t)$ (in their notation), which is the potential difference between the solid-phase potential evaluated at the positive and negative current collectors. We also note that in \cite{kemper2013extended} current has been defined in the direction of flow of positive charge (opposite of the convention used here) so we account for the sign change here. The model in \cite{kemper2013extended} is then taken to be:
\begin{align}
    V =& U_{\text{eq}} + \eta\ts{r} + \eta\ts{c} + \Delta\Phi\ts{Elec},
\end{align} 
\begin{align}
     U\ts{eq} =& U\ts{p}(c\ts{s,p}^0\big|_{r=1})-U\ts{n}(c\ts{s,n}^0\big|_{r=1}), \\
     \begin{split}
	   \eta\ts{r} = & - 2\sinh^{-1}\left(\frac{I}{L\ts{p} j\ts{0,p}(1)}\right) 
        - 2\sinh^{-1}\left(\frac{I}{L\ts{n} j\ts{0,n}(0)}\right) \\ 
					& + \frac{2 \mathcal{C}\ts{e}}{\gamma\ts{e}} \frac{\sigma\ts{p}}{\mybar{\kappa}\ts{e,p} + \frac{\mathcal{C}\ts{e}}{\gamma\ts{e}}\sigma\ts{p}}\left(\sinh^{-1}\left(\frac{I}{L\ts{p} j\ts{0,p}(1)}\right) - \sinh^{-1}\left(\frac{I}{L\ts{p} j\ts{0,p}(1-L\ts{p})}\right) \right) \\ 
					& + \frac{2 \mathcal{C}\ts{e}}{\gamma\ts{e}} \frac{\sigma\ts{n}}{\mybar{\kappa}\ts{e,n} + \frac{\mathcal{C}\ts{e}}{\gamma\ts{e}}\sigma\ts{n}}\left(\sinh^{-1}\left(\frac{I}{L\ts{n} j\ts{0,n}(0)}\right) - \sinh^{-1}\left(\frac{I}{L\ts{n} j\ts{0,n}(L\ts{n})}\right) \right), 
        \end{split}\\ 
        \begin{split}
        \eta_c =& 2(1-t^+)\left( 
		  \frac{\mybar{\kappa}\ts{e,p}}{\mybar{\kappa}\ts{e,p}+\frac{\mathcal{C}\ts{e}}{\gamma\ts{e}}\sigma\ts{p}}\log\left(\frac{1+\mathcal{C}\ts{e} c\ts{e,p}^1\big|_{x=1}}{1+\mathcal{C}\ts{e} c\ts{e,p}^1\big|_{x=1-L\ts{p}}}\right)
        + \log\left( \frac{1+\mathcal{C}\ts{e} c\ts{e,s}^1\big|_{x=L\ts{n}}}{1+\mathcal{C}\ts{e} c\ts{e,s}^1\big|_{x=1-L\ts{p}}} \right) \right. \\
        & \hphantom{2(1-t^+) (1-t)}
		\left. +\frac{\mybar{\kappa}\ts{e,n}}{\mybar{\kappa}\ts{e,n}+\frac{\mathcal{C}\ts{e}}{\gamma\ts{e}}\sigma\ts{n}}\log\left(\frac{1+\mathcal{C}\ts{e} c\ts{e,n}^1\big|_{x=L\ts{n}}}{1+\mathcal{C}\ts{e} c\ts{e,n}^1\big|_{x=0}}\right)
        \right),
        \end{split} \\
	 \Delta\Phi_{\text{Elec}} =& - \frac{I}{\hat{\kappa}\ts{e}}\left( \frac{L\ts{n}}{\mybar{\kappa}\ts{e,n} + \mathcal{C}\ts{e} \gamma\ts{e}\sigma_n}
        - \frac{L_s}{\mybar{\kappa}\ts{e,s}} + \frac{L\ts{p}}{\mybar{\kappa}\ts{e,p}+\mathcal{C}\ts{e} \gamma\ts{e} \sigma_p}\right),
\end{align}
where $\mybar{\kappa}\ts{e,k}$ is given by (\ref{eqn:sigma_bar_notation}) and 
\begin{equation*}
    j\ts{$0$,k} = \frac{\gamma\ts{k}}{\mathcal{C}\ts{r,k}} (c\ts{s,k}^0)^{1/2}(1-c\ts{s,k}^0)^{1/2}(1+\mathcal{C}\ts{e} c\ts{e,k}^1)^{1/2}.
\end{equation*}

\newpage

\section{Tables}

\begin{table}[h]
	\centering 
	\resizebox{\textwidth}{!}{%
	\begin{tabular}{c c l c c c} 
	\toprule
     Parameter & Units & Description & $\Omega^*\ts{n}$ & $\Omega^*\ts{s}$ & $\Omega^*\ts{p}$ \\ 
    \midrule 
    $\epsilon\ts{k}$ & - & Electrolyte volume fraction & $0.3$ & 1 & $0.3$  \\
    $c^*\ts{k,max}$ & $\SI{}{\mol / \metre^{3}}$ & Maximum lithium concentration &$2.4983\times 10^4$ & - & $5.1218\times 10^4$ \\
     $\sigma^*\ts{k}$ & $\SI{}{ \siemens / \metre}$ & Solid conductivity & 100 & - & 10 \\
     $D^*\ts{s,k}$ & $\SI{}{\metre^{2}/\second}$ & Electrode diffusivity &  $3.9\times10^{-14}$ & - &  $1\times10^{-13}$\\
     $R^*\ts{k}$ & $\SI{}{\micro\metre}$ & Particle radius & 10 & - & 10  \\
     $a^*\ts{k}$ & $\SI{}{\micro\metre^{-1}}$ & Electrode surface area density & 0.18 & - & 0.15 \\
     $m^*\ts{k}$ & $\SI{}{(\ampere / \metre^{2})(\metre^3/\mol)^{1.5}}$ & Reaction rate & $2\times10^{-5}$ & - & $6\times10^{-7}$ \\
     $L^*\ts{k}$ & $\SI{}{\micro\metre}$ & Thickness & 100 & 25 & 100 \\
     $U^*\ts{k,ref}$ & $\SI{}{\volt}$ & Reference OCP& 0.18 & - & 3.94 \\
      \midrule
    $c^*\ts{e, typ}$ & $\SI{}{\mol / \metre^{3}}$ & Typical lithium-ion concentration in electrolyte & & $1\times10^3$ & \\
    $D\ts{e,typ}^*$ & $\SI{}{\metre^{2}/\second}$ & Typical electrolyte diffusivity &  & $5.34\times10^{-10}$ &  \\
    $\kappa\ts{e,typ}^*$ & $\SI{}{\siemens/\metre}$ & Typical electrolyte conductivity &  & $1.1$ &  \\
     $F^*$ & $\SI{}{\coulomb / \mol}$ & Faraday's constant & & 96485 &  \\
     $R^*$ & $\SI{}{\joule / (\mol \, \kelvin)}$ & Universal gas constant && 8.314472 & \\
     $T^*$ & $\SI{}{\kelvin}$ & Temperature && 298.15 & \\
     $b$ & - & Bruggeman coefficient && 1.5 &  \\
     $t^+$ & - & Transference number && 0.4 & \\
     $I^*\ts{typ}$ & $\SI{}{\ampere / \metre^2}$ & Typical current density & & 24 ($\SI{1}{C}$)& \\
    \bottomrule
	\end{tabular}}
    \caption{Dimensional model parameters with values taken from \cite{SMouraGithub}.} \label{tab:parameters}
\end{table}

\begin{table}[h]
	\centering 
	\resizebox{\textwidth}{!}{%
	\begin{tabular}{c c l c} 
	\toprule
     Symbol & Expression & Interpretation & Value $[\SI{}{s}]$  \\ 
    \midrule 
    $\tau\ts{d}^*$ & ${F^* c^*\ts{n,max} L^*}/{I\ts{typ}^*}$ & Discharge timescale &  $2.2598 \times 10^4 / \mathcal{C}$ \\
    $\tau\ts{n}^*$ & ${(R\ts{n}^*)^2}/{D^*\ts{s,p}}$ & Diffusion timescale in the negative electrode particle & $2.5641 \times 10^3$ \\
    $\tau^*\ts{p}$ & ${(R^*\ts{p})^2}/{D^*\ts{s,p}}$ & Diffusion timescale in the positive electrode particle & $1 \times 10^3$ \\
    $\tau^*\ts{e}$ & ${(L^*)^2}/{D^*\ts{e,typ}}$ & Diffusion timescale in the electrolyte & 94.803 \\
    $\tau\ts{r,n}^*$ & ${F^*}/({m^*\ts{n} a^*\ts{n} (c^*\ts{e,typ})^{1/2}})$ & Reaction timescale in the negative electrode & 847.534\\
    $\tau\ts{r,p}^*$ & ${F^*}/({m^*\ts{p} a^*\ts{p} (c^*\ts{e,typ})^{1/2}})$ & Reaction timescale in the positive electrode & $3.3901 \times 10^4$\\
    \bottomrule          
    \end{tabular}}
    \caption{Timescales associated with the physical processes occurring within the battery model. Here, $\mathcal{C}$ is the C-rate.}
    \label{tab:timescales}
\end{table}

\begin{table}[h]
	\centering 
	\resizebox{\textwidth}{!}{%
	\begin{tabular}{c c p{7cm} c c c c c} 
	\toprule
     Parameter & Expression & Interpretation & n & s & p  \\ 
    \midrule 
    $L\ts{k}$ & $L\ts{k}^*/L^*$ & Ratio of region thickness to cell thickness & 0.4444 & 0.1111 & 0.4444 \\ 
    $\mathcal{C}\ts{k}$ & $\tau\ts{k}^*/\tau\ts{d}^*$ & Ratio of solid diffusion and discharge timescales & 0.1134 $\mathcal{C}$ & - & 0.0442 $\mathcal{C}$ \\ 
    $\mathcal{C}\ts{r,k}$ & $\tau\ts{r,k}^*/\tau\ts{d}^*$ & Ratio of reaction and discharge timescales& 0.0375 $\mathcal{C}$ & - & 1.5 $\mathcal{C}$\\     
   $\sigma\ts{k}$ & $(R^* T^* / F^*) / ( (I\ts{typ}^* L^* /\sigma\ts{k}^*))$ & Ratio of thermal voltage and typical Ohmic drop in the solid & $475.791/ \mathcal{C}$ & - & $47.5791/ \mathcal{C}$ \\ 
    $a\ts{k}$ & $a\ts{k}^* R\ts{k}^*$ & Product of particle radius and surface area density   & 1.8 & - & 1.5 \\ 
   $\gamma\ts{k}$ & $c\ts{k,max}^*/c\ts{n,max}^*$ & Ratio of maximum lithium concentrations in solid & 1  & - & 2.0501\\
\midrule
    $\mathcal{C}\ts{e}$ & $\tau\ts{e}^*/\tau\ts{d}^*$ & Ratio of electrolyte transport and discharge timescales  && $4.19\times10^{-3} \: \mathcal{C}$ & \\ 
    $\gamma\ts{e}$ & $c\ts{e,typ}^* / c\ts{n,max}^*$ & Ratio of maximum lithium concentration in the negative electrode solid and typical electrolyte concentration && $0.04$ & \\
    $\hat{\kappa}\ts{e}$ & $\left(R^* T^* / F^* \right) / \left( (I\ts{typ}^* L^* / \kappa\ts{e,typ}^*) \right) $ & Ratio of thermal voltage to the typical Ohmic drop in the electrolyte & & $5.2337 / \mathcal{C}$ & \\
    \midrule 
    $c\ts{k,0}$ & $c\ts{k,0}^*/c\ts{k,max}^*$ & Ratio of initial lithium concentration to maximum lithium concentration in solid & 0.8 & - & 0.6 \\
    \bottomrule          
    \end{tabular}}
    \caption{Typical dimensionless parameter values. Here $\mathcal{C}=I^*/(\SI{24}{\ampere \metre^{-2}})$ is the C-Rate where we have taken a 1C rate to correspond to a typical $x$-direction current density of \SI{24}{\ampere \metre^{-2}}. This is for a cell with an initial stoichiometry of $0.8$ in the negative electrode and $0.6$ in the positive electrode with a voltage cutoff of \SI{3.2}{\volt}.}
    \label{table:dimensionless_parameter_values}
\end{table}

\begin{table}[h]
  \centering
  \begin{tabular}{c c c c c c}
    \toprule
    & \SI{0.1}{C} & \SI{0.5}{C} & \SI{1}{C} & \SI{2}{C} & \SI{3}{C} \\
    \midrule
    SPM & \SI{1.72}{mV} & \SI{9.62}{mV} & \SI{19.86}{mV} & \SI{40.67}{mV} & \SI{62.78}{mV} \\
    SPMe & \SI{0.17}{mV} & \SI{1.34}{mV} & \SI{3.04}{mV} & \SI{7.36}{mV} & \SI{13.34}{mV} \\
    \bottomrule
  \end{tabular}
  \caption{RMS voltage error between the reduced models and the DFN model for the finite-volume implementation with 30 points in each domain, as discussed in Section~\ref{sec:SPM-SPMe-DFN}.}
  \label{tab:RMSE-table}
\end{table}

\begin{table}[h]
  \centering
  \begin{tabular}{c c c c c}
    \toprule
            & 5~pts & 10~pts & 20~pts & 30~pts \\
    \midrule
      SPM   & 0.24~s & 0.25~s  & 0.34~s  & 0.41~s \\
      SPMe  & 0.50~s & 0.58~s & 0.56~s & 0.59~s \\
      DFN   & 3.34~s & 5.10~s & 18.51~s & 61.36~s \\
    \bottomrule
  \end{tabular}
    \caption{Average computation time across C-rates (0.1~C, 0.5~C, 1~C, 2~C, 3~C) for the SPM (\ref{eqn:SPM}), SPMe (\ref{eqn:SPMe}), and DFN model (\ref{eqn:DFN}). Here,`$n$~pts' refers to $n$ points in each domain.}
\end{table}

\begin{table}[h]
	\centering 
	\begin{tabular}{c l p{5cm}} 
  \toprule
  Parameter combination & Required size & Interpretation\\
  \midrule
  $\mathcal{C}\ts{e} = I\ts{typ}^* L^*/(D\ts{e,typ}^*F^*c\ts{n,max}^*)$  & $\ll 1$ &Lithium-ion migration timescale is small relative to typical discharge timescale \\
  $R^* T^* \sigma\ts{k}^* / (F^* I\ts{typ}^* L^*)$  & $\gg 1$ & Thermal voltage is large relative to the typical potential drop in electrode k\\
  $R^* T^* \kappa\ts{e,typ}^*/(F^* I\ts{typ}^* L^*)$ & $\gg 1$ & Thermal voltage is large relative to the typical potential drop in the electrolyte \\
  $(R\ts{k}^*)^2 I\ts{typ}^*/(D\ts{s,k}^* F^* c\ts{n,max}^* L^*)$ & $\ll 1/\mathcal{C}\ts{e}$ & Solid diffusion occurs on a shorter or similar timescale to a discharge  \\
  $I\ts{typ}^*/(m\ts{k}^* a\ts{k}^* (c\ts{e,typ}^*)^{1/2} c\ts{n,max}^* L^*)$ 
  & $\ll 1/\mathcal{C}\ts{e}$ & Reactions occur on a shorter or similar timescale to a discharge \\
  \bottomrule
  \end{tabular}
  \caption{The key conditions to be satisfied for the application of (\ref{eqn:SPMe-dimensional}). In addition, it is required that $\mathcal{C}\ts{e} \ll L\ts{k}^*/L^* \ll 1/\mathcal{C}\ts{e}$, $\mathcal{C}\ts{e}\ll c\ts{p,max}/c\ts{n,max}\ll 1/\mathcal{C}\ts{e}$, and $\mathcal{C}\ts{e} \ll c\ts{e,typ}/c\ts{n,max}\ll 1/\mathcal{C}\ts{e}$, which are true in practical situations.}
  \label{tab:conditions}
\end{table}

\newpage

\section*{Acknowledgements}
This publication is based on work supported by the EPSRC Centre For Doctoral Training in Industrially Focused Mathematical Modelling (EP/L015803/1) in collaboration with Siemens Corporate Technology and BBOXX. This work was also supported with funding provided by The Faraday Institution, grant number EP/S003053/1, FIRG003.

\end{document}